\documentclass[manuscript,screen,authorversion]{acmart}

\begin{document}

\title{Decolonial AI Alignment: Openness, Vi\'{s}e\d{s}a-Dharma, and Including Excluded Knowledges}

\author{Kush R. Varshney}
\email{krvarshn@us.ibm.com}
\orcid{0000-0002-7376-5536}
\affiliation{%
  \institution{IBM Research -- Thomas J. Watson Research Center}
  \streetaddress{1101 Kitchawan Road}
  \city{Yorktown Heights}
  \state{New York}
  \country{USA}
  \postcode{10598}
}


\begin{abstract}
Prior work has explicated the coloniality of artificial intelligence (AI) development and deployment through mechanisms such as extractivism, automation, sociological essentialism, surveillance, and containment. However, that work has not engaged much with alignment: teaching behaviors to a large language model (LLM) in line with desired values, and has not considered a mechanism that arises within that process: moral absolutism---a part of the coloniality of knowledge. Colonialism has a history of altering the beliefs and values of colonized peoples; in this paper, I argue that this history is recapitulated in current LLM alignment practices and technologies. Furthermore, I suggest that AI alignment be decolonialized using three forms of openness: openness of models, openness to society, and openness to excluded knowledges. This suggested approach to decolonial AI alignment uses ideas from the argumentative moral philosophical tradition of Hinduism, which has been described as an open-source religion. One concept used is vi\'{s}e\d{s}a-dharma, or particular context-specific notions of right and wrong. At the end of the paper, I provide a suggested reference architecture to work toward the proposed framework.
\end{abstract}




\maketitle

\section{Introduction}
\label{sec:intro}

For more than a year now, the public has experienced powerful large language models (LLMs) such as GPT-4, Claude 2, Gemini, Llama 3, and Mixtral. Beyond the initial amazement and excitement, we have witnessed the bearing out of environmental and sociotechnical harms foreseen by  Ref.\ \cite{BenderGMS2021} and others. The need to control the cost and behavior of LLMs has become apparent. While such governance is relevant in chat interfaces made available by model providers, it comes to the forefront when LLMs are infused into software applications and use cases by organizations with varied affected communities, missions, goals, regulations, and values.  

The way in which an LLM may be infused into an application, and the degree to which it may be customized \cite{KirkVRH2023}, depends on what the model provider allows. Despite the term `open' being used and abused in different ways by model providers \cite{widder2023open}, these are questions of openness. A provider may only allow application programming interface (API) access to a closed proprietary model. A provider may license model weights and parameters to users so that they may download the LLM locally and fine-tune it on their own data. A provider may offer full transparency into the pre-training datasets, data pre-processing operations, and architecture that would allow others to recreate their model. Although subject to `open-washing,' we are seeing an emerging divergence between advocates of `open' vs.\ `closed' LLMs in the market, epitomized by the AI Alliance vs.\ the Frontier Model Forum \cite{alliance2023}.

The more open LLMs are, the more they permit application developers to make them authentic to their needs and the values of their communities. For example, Jacaranda Health has created UlizaLlama\footnote{\url{https://huggingface.co/Jacaranda/UlizaLlama}} for its community in East Africa by continuing to train Llama 2 with 322M tokens of Kiswahili and further instruction fine-tuning it to respond to questions in healthcare, agriculture, and other locally-relevant topics. UlizaLlama is a step in Jacaranda's development of its LLM-infused maternal health digital platform PROMPTS. In constrast, application developers and their communities are \emph{not} empowered to reflect their own values with \emph{closed} LLMs. They must live with the commandments of good and bad, and right and wrong that the provider of a closed model happens to have inserted.

The further actions to change an LLM's behavior, beyond the standard pre-training of a base LLM, have come to be known as \emph{alignment}. The term is an empty signifier without a fixed concept that is signified; different parties have appropriated the term to refer to various actions for getting an LLM to behave according to some human values \cite{kirk2023empty}. Desired behaviors (with varying levels of specificity) could range from following instructions, to carrying on helpful conversations, to yielding safe or moral outputs (with different definitions), to something else altogether. The behavior of an LLM may be controlled through data curation, full or parameter-efficient supervised fine-tuning, reinforcement learning with direct or indirect human feedback, self-alignment, prompt engineering (few-shot learning), and guardrails or moderations \cite{ji2023ai,wang2023aligning,kirk-etal-2023-past}. I defer discussion of the details, and of the computation, data and human resources required for each of these approaches to Section \ref{sec:prelim:llm}. Existing approaches do not allow for different aligned behaviors of a given LLM based on the context of deployment. 

A further question with LLM-infused applications has to do with the business notion of `value,' as in earnings, profits, or other measures of commercial utility, rather than human values of right and wrong. Does value accrue to the model provider or to the application developer \cite{BornsteinAC2023}? Openness may enable ecosystems in which application developers and their communities accrue value (cf.\ Jacaranda Health), whereas closed LLM providers may exercise their power to be extractive in nature. Extractivism is a part of \emph{coloniality} \cite{ricaurte2019data}, which is the main topic of this paper.

Colonialism is one country controlling another and exploiting it economically and in other ways. Coloniality, however, describes domination, including in abstract forms such as in the production of knowledge, that remains after the end of formal colonialism \cite{quijano2007coloniality}. \emph{Decoloniality} is the process of challenging and dismantling coloniality \cite{mignolo2010introduction}. The terms usually refer to European or Western colonialism and its remnants in the Global South. Decolonial \emph{computing} is developing computing systems with and for people there that reduce asymmetric power relationships, based on their values and their knowledge systems \cite{ali2016brief}. Based on these ideas, there has been a recent flowering of research on decolonial artificial intelligence (AI), beginning with the seminal paper by Mohamed, Png and Isaac \cite{mohamed2020decolonial}. Through this lens, extractive providers of closed models may be viewed as \emph{metropoles}: the colonial powers. Further discussion of the decolonial AI literature is provided in Section \ref{sec:prelim:colonial}.

The scope of the coloniality considered in AI thus far has included extractivism as well as four other mechanisms: automation, sociological essentialism, surveillance, and containment \cite{tacheva2023ai}. The contribution of this paper is to examine a different colonial mechanism from these five, namely \emph{ethical essentialism} also known as \emph{moral absolutism}, which arises specifically in the \emph{alignment} of LLMs. If a powerful model provider views their (Western) ethics or moral philosophy as universally correct, leaves no possibility for moral variety \cite{flanagan2016geography}, and marginalizes all other ways of thinking about right and wrong, then their approach to AI alignment is colonial. They are behaving as a metropole.

In Section \ref{sec:colonial-align}, I expand upon this viewpoint of coloniality occurring in AI alignment through the mechanism of moral absolutism and the centering of Western philosophy. This includes not only a philosophical discussion, but also a critical examination of the technology for AI alignment. In Section \ref{sec:decolonial}, I broach the decolonialization of AI alignment, which can be seen as a kind of decolonialization of knowledge, through the lens of open science and innovation \cite{chan2020open}. Such openness includes three thrusts: (1) openness to research artifacts (which includes LLMs in our context), (2) openness to society, and (3) openness to excluded knowledges \cite{chan2020open}. Based on these three openness pursuits, I lay out three desiderata for doing AI alignment in a decolonial manner. Furthermore, I suggest an approach to alignment that meets the desiderata. This suggested approach builds upon the non-universal non-absolutist tradition of moral philosophy known as Hinduism \cite{dhand2002dharma,ranganathan2022hinduism}, which includes vibrant argument and debate on the nature of \emph{dharma} (right behavior) and its explication through various ways of knowing, including artistic expression \cite{divakaran2023broadening}. The syncretic framework of Hinduism (described in greater detail in Section \ref{sec:prelim:hindu}) has the appropriate characteristics of openness to be used as a starting point for an alternative future of AI alignment \cite{siddhartha2008open,Schrei2010}. At the end, I build upon the suggested dharmic approach and give a more concrete reference architecture of a technology stack for less morally absolute and less colonialized AI alignment. 

\section{Preliminaries}
\label{sec:prelim}

\subsection{Large Language Model Development Lifecycle and Alignment}
\label{sec:prelim:llm}

The currently prevalent development lifecycle for applications infused with LLMs may be divided into two halves: steps carried out by model providers and steps carried out by application developers. In an \emph{imperfect} analogy with teaching a child, the model provider does the basic steps of teaching the LLM to go from babbling words, to having fluency in language, to following instructions, to carrying on a conversation. The application developer, if so empowered, teaches the LLM \emph{culture}, which may include steps on subject matter expertise, social norms, laws, customs, and beliefs. Getting to the point of language fluency may be termed pre-training the base model or foundation model. Any of the steps after language fluency may be called `alignment,' depending on the interlocutor. As mentioned earlier, the term `alignment' is an empty signifier, so it is not fixed to refer to any specific step \cite{kirk2023empty}.

In pre-training, some amount of enculturation is possible by curating the content of the training dataset to include an abundance of topics that the model provider wishes the LLM to be skilled in and filtering out taboo topics. As discussed further in Section \ref{sec:prelim:colonial:ai}, some amount of undesirable cultural knowledge leaks into the pre-training performed by the model provider. Filtering is computationally-intensive given the size of datasets being in the trillions of tokens. Data curation is followed by self-supervised learning (like a peekaboo game) to obtain the base model, which may take months despite using thousands of high-end graphical processing units. 

The AI technologies to do any of the alignment steps on top of the base model are essentially the same, whether the goal is following instructions, behaving according to social norms, or something else. Several techniques exist with varying resource requirements for humans, data, and computation. Supervised fine-tuning (SFT) updates all of the model weights according to a smaller, but still large dataset containing data with both inputs and outputs. It is fairly computationally-intensive given that all weights are updated. If a model has already been trained to follow instructions, then a dataset with instructions, inputs, and outputs may be used. To reduce data and computational complexity, parameter-efficient fine-tuning methods do not update all model weights, but are more frugal. One specific approach, low-rank adaptation (LoRA), trains a matrix of weights of the same dimensions as the LLM weights. This LoRA matrix is added to the LLM weights at the time of inference. However, the LoRA matrix has orders of magnitude fewer degrees of freedom through its construction as a low-rank matrix and is thus more efficient \cite{hu2021lora}.

Several alignment techniques include full SFT as a module, including reinforcement learning from human feedback (RLHF) \cite{Ouyang2022}, reinforcement learning from AI feedback (RLAIF) \cite{Bai2022}, and self-align \cite{sun2023principle}. After SFT, these methods further align the LLM by feeding back judgements of which outputs are preferred by, respectively, either: humans, a preference model trained according to a set of explicit regulations (a constitution), or an LLM prompted through instructions to respect a set of explicit regulations. RLHF requires a large amount of human labor and all three are computationally involved. The LLM prompted to respect a set of explicit regulations in the self-align approach is also, by itself, a simple but not always reliable way to align a model. By manually designing system prompts or prompt templates to accompany all inputs, the LLM's behavior may be controlled. Such prompt engineering adds to inference costs because the input to the model includes extra tokens every time.

Finally, another way to align the behavior of an LLM is by a post-processing module that examines the output and determines whether it satisfies pre-determined guardrails for specific unwanted behaviors. These post-processors or moderations may be small classifiers or other LLMs acting as `judges' \cite{zheng2023judging,achintalwar2024detectors}.

\subsection{Coloniality and Decoloniality}
\label{sec:prelim:colonial}

As introduced in Section \ref{sec:intro}, coloniality is an extension of colonialism after its formal end. It is the values, ways of knowing, and power structures instituted during colonialism that remain, and may even be expanded to places without a history of colonialism, that rationalize and perpetuate Western dominance. Decolonial perspectives disobey the program of coloniality \cite{mignolo2018decoloniality}. The theory of coloniality includes coloniality of power, coloniality of knowledge, and coloniality of being. Coloniality of power describes social discrimination through hierarchies and caste systems instituted during colonialism \cite{quijano2007coloniality}. Coloniality of knowledge is the suppression of colonized peoples' culture and ways of knowing; it is used by colonizers in service of coloniality of power \cite{quijano2007coloniality}. Coloniality of being is a severe version of coloniality of knowledge: a people's knowledge system is so inferior that those people do not even deserve to \emph{be}, or to be human \cite{maldonado2007coloniality}. Coloniality is a subset of Empire \cite{hardt2001empire}, which also includes other dimensions of hegemony such as heteropatriarchy and white supremacy \cite{tacheva2023ai}. Coloniality and decoloniality influence several areas of study, including international relations, development theory, communication theory, human-computer interaction, and many others \cite{zondi2020decolonising,patel2020race,na2022mapping,alvarado2021decolonial,pendse2022treatment}. 

\subsubsection{Excluded Knowledges}
\label{sec:prelim:colonial:knowl}

The main aspect of the coloniality of knowledge is the imposition of Western epistemologies, or ways of knowing, and the suppression of non-Western epistemologies. This suppression is often a violent extermination of a knowledge system termed epistemicide \cite{de2011epistemologias,grosfoguel2013structure}. The \emph{excluded knowledges} of the colonized or marginalized groups may come from ``organic, spiritual and land-based systems'' or arise from social movements \cite{hall2017decolonization}. As Hlabanganeh explains \cite{hlabangane2021underside}: ``These other ways of knowing and being are rendered unintelligible when filtered through Western sensibilities that, for example, set greater store by the mind in juxtaposition with and preference to the body and spirit, that prioritise instrumental/rational pursuits such as profit which lead to individualism, and that conceive of nature and culture as dichotomous entities with culture gaining mastery over nature. While these ways of being and knowing have been exalted to represent the epitome of evolution, so to speak, they are in fact particular to a certain way of thinking.''

Hall and Tandon's decolonial knowledge democracy acknowledges these multiple epistemologies and recognizes that knowledge comes in many forms beyond natural language text, including images, music, drama, ceremony, and meditation \cite{hall2017decolonization}. It sees open access and sharing of this knowledge as a means for decolonialization \cite{hall2017decolonization}. Decolonializing knowledge is often done by making teaching materials, curricula, practices, and institutions more open and inclusive \cite{patin2021interrupting}. Referring back to the teaching analogy of LLM alignment in Section \ref{sec:prelim:llm}, we will later see how the proposed decolonial AI alignment also makes teaching more open and inclusive.

\subsubsection{Coloniality and Moral Philosophy}
\label{sec:prelim:colonial:moral}

Within the coloniality of knowledge is knowledge systems of values. Values are the realm of moral philosophy, the branch of philosophy that studies right and wrong \cite{ErmanM2018}. Historically, colonialism altered the beliefs and values of colonized peoples. For example, Igboin writes \cite{igboin2011colonialism}: ``Colonial rule disrupted the traditional machinery of moral homogeneity and practice. The method of moral inculcation was vitiated, which resulted in the abandonment of traditional norms and values through a systematic depersonalisation of the African and paganisation of its values. Instead of the cherished communalism which defined the life of the African, for example, a burgeoning societal construct was introduced which alienates and destroys the organic fabric of the spirit of we-feeling.'' On Ranganathan's account, during and after the Western colonization of India, ``Hindus adopted a West-centric frame for understanding their tradition as religious because of colonization'' \cite{ranganathan2022hinduism}. This phenomenon was not merely a side-effect, but a goal of the program of colonialism. ``For Western colonialism to succeed, philosophy and explication---South Asian moral philosophy---has to be erased, as it constitutes a critical arena for the West's claim to authority'' \cite{ranganathan2022hinduism}. The colonizers positioned their Western philosophical tradition as rational and secular, and the default; they erased the Hindu traditions as the irrational, unjustified `other.'

The erasure of systems of morality extends from colonialism to coloniality \cite{dunford2017toward,van2020studying,motilal2021care}. Maldonado-Torres emphasizes that \cite{maldonado2017religion}: ``The concept of religion most used in the West by scholars and laypeople alike is a specifically modern concept forged in the context of imperialism and colonial expansion.'' This concept includes the idea that a religion must have a single book as its authority.

\subsubsection{Coloniality and Artificial Intelligence}
\label{sec:prelim:colonial:ai}

Within decolonial computing \cite{ali2016brief} is the study of AI. AI is value-laden; the term itself reflects the legacy of dominance hierarchies such as man over nature, patriarchy, colonialism, and racism \cite{cave2020problem}. Now in the age of powerful LLMs, historical dominance is getting even more entrenched. For example, empirical analysis shows that LLMs have sociopolitical biases in favor of dominant groups \cite{Feng2023,Durmus2023}. They exhibit West-centric biases in representing moral values \cite{BenklerMFSS2023}. In addition, morality captured by multi-lingual language models does not reflect cultural differences, but rather is dominated by high-resource languages and cultures \cite{haemmerl-etal-2023-speaking}. 

When researchers and activists were first sounding the alarm that LLMs would harm marginalized communities by encoding and reinforcing hegemonic viewpoints, the charge of hegemony rested on unfathomably large training datasets scraped from the bottom of the barrel of the internet that over-represent white supremacist, misogynist, and ageist content \cite{BenderGMS2021}. However, it has now become apparent that the behavior of performant LLMs depends as much on their alignment as on the training data \cite{Ouyang2022,Bai2022,Wu2023,casper2023open}. The workers laboring to give human feedback for alignment, often located in poor communities, may be traumatized and scarred \cite{Perrigo2023,Kantrowitz2023}. Although there are exceptional examples of workers and communities being uplifted \cite{Mehrotra2022,Perrigo2023b}, the process usually recapitulates exploitation colonialism: a small number of powerful companies using the workers to increase their own power and wealth while little benefit and an abundance of negative externalities are left in the workers' communities \cite{e53927b8-0a79-3d19-93b7-95642d44b953}. 

Research at the intersection of AI and coloniality is not new. The seminal work by Mohamed, Png and Isaac \cite{mohamed2020decolonial}, a series of articles in MIT Technology Review by Hao et al.~\cite{Hao2022}, and other prior work \cite{Manyfesto,birhane2020algorithmic,costanza2020design,adams2021can,crawford2021atlas,ricaurte2022ethics,ehsan2022algorithmic,hassan2023governing,muldoon2023artificial} is focused on five mechanisms taxonomized by Tacheva and Ramasubramanian \cite{tacheva2023ai}: (1) extractivism, (2) automation, (3) sociological essentialism, (4) surveillance, and (5) containment. Extractivism entails the extraction of labor, materials, and data, including the human feedback mentioned above, and datafication that extracts the digital breadcrumbs of people to be bought and sold. The tenor of AI for social good---bestowing technology on the underdeveloped---may also be extractive if it leads to corporate capture \cite{green2019good,viera2021giving}. Automation involves the replacement of (empathetic) human decision making with biased machine decision making in consequential domains that especially hurts members of minoritized groups \cite{d2022empathy,knowles2023trustworthy} as well as `ghost work' and `fauxtomation' that present a veneer of objectivity, but actually involve people behind the scenes exploited as a digital underclass \cite{gray2019ghost}. Sociological essentialism erases the nuance behind different identities and cultures through the use of broad categories \cite{buolamwini2018gender,benthall2019racial}. AI-based surveillance, including biometric mass surveillance, is especially hurtful to people facing power asymmetries \cite{benjamin2022viral}. Containment,  technological apartheid, digital redlining, and censorship involve the powerful using AI technologies to police who belongs where \cite{adebisi2014knowledge,lambright2019digital}. As discussed in Section \ref{sec:prelim:colonial:moral}, the coloniality of knowledge may include the erasure of knowledge systems of ethics, moral philosophy, and reasoning about values. The existing work on decolonial AI described thus far has not focused on morality. Thus, a sixth mechanism for colonial AI, beyond the five in Tacheva and Ramasubramanian's taxonomy, is emerging alongside the emergence of LLM alignment: ethical essentialism or moral absolutism. 

\subsection{Hinduism and Dharma}
\label{sec:prelim:hindu}

Hinduism is the name applied by outsiders to the multifarious collection of moral philosophies originating in the Indian subcontinent. It is a religion without a single founder, book, dogma, or set of practices. 

\subsubsection{Basic Concepts and Openness}
\label{sec:prelim:hindu:basic}

The main concept of Hinduism is brahman, a force or ultimate reality that pervades the universe; xe is described as sat-cit-\={a}nanda or truth-consciousness-bliss \cite{Maheshwarachary1988,dhand2002dharma,tharoor2018hindu}. The universe is made up of \={a}tman---the essence of each individual that persists across lifetimes---and prak\d{r}ti---solid, liquid, gas, energy, and space. The \={a}tman wanders through cycles of birth, life, and death---sa\d{m}s\={a}ra---with the aim of attaining mok\d{s}a: freedom from sa\d{m}s\={a}ra and union with brahman. \emph{Dharma} consists of the notions of righteousness and moral values appropriate for the \={a}tman. Following dharma helps the \={a}tman advance toward mok\d{s}a.

As mentioned above, Hinduism is not dogmatic, doctrinaire, or morally absolutist. Commentators have described it as \emph{open-source} \cite{siddhartha2008open,Schrei2010}. The kernel is the Vedas, a set of scriptures that includes the idea `eka\d{m} sat vipr\={a} bahudh\={a} vadanti': there are many wise ways to reach the one truth, to reach brahman (\d{R}g Veda, mandala 1, hymn 164, verse 46). As such, there are hundreds of thousands of additional scriptures and philosophies that extend, fork, fine-tune, and contradict themselves and the Vedas. Shani and Chadha Behera explain that \cite{shani2022provincialising}: ``the concept of dharma offers a mode of understanding the multidimensionality of human existence without negating any of its varied, contradictory expressions.'' For example, C\={a}rv\={a}ka, Buddhist, Jain, and other so-called n\={a}stika samprad\={a}yas (knowledge systems) reject the Vedas.\footnote{C\={a}rv\={a}ka philosophy is nihilistic and rejects much more than just the Vedas.} Moreover, even within \={a}stika samprad\={a}yas that accept the Vedas, their utility is questioned. For example, the Bhagavad-G\={\i}t\={a} says that the Vedas are of limited use to people who have understood their main message (chapter 2, verse 46). Such `heresy' is not only tolerated, it is accepted and encouraged.\footnote{Bhagavad-G\={\i}t\={a}, chapter 18, verse 63 and R\={a}ma-Carita-M\={a}nasa, book 7, verse 42 encourage the follower to do as they see fit.}

The knowledge systems and scriptures referred to above are expressed in many forms, including the Vedas (sacred utterances, descriptions of rituals, and their explanations), Upani\d{s}ads (discussions of meditation, philosophy, consciousness, and ontological knowledge), \'{s}\={a}stras (treatises on law, architecture, astronomy, etc.), itih\={a}sas (epics), pur\={a}\d{n}as (lore), and dar\'{s}anas (philosophical literature on spirituality). The different literatures are directed toward different people: some more popular and others more scholarly. Different paths to mok\d{s}a, including devotion, work, and knowledge, are directed toward different people depending on their characteristics. For example, myriad gods and goddesses representing different aspects of brahman are available to devotees depending on their wishes. Morality is primarily presented by example or metaphor through stories in itih\={a}sas and  pur\={a}\d{n}as (including renditions in drawing, sculpture, dance, etc. \cite{divakaran2023broadening}) rather than by explicit commandment in treatises \cite{dhand2002dharma}.

\subsubsection{Vi\'{s}e\d{s}a-Dharma}
\label{sec:prelim:hindu:vises}

Unlike the goal of finding universally-applicable moral philosophies presupposed in the West,\footnote{Some traditions that were colonized by the West also aim for universal theories. The general trend in Western ethics is toward universality, but there are exceptions, cf.\ Bernard Williams.} there is no desire to identify universal ethical principles in Hinduism \cite{dhand2002dharma}. \emph{Dharma} was richly debated in pre-colonial India. There were deontological philosophies (e.g.\ m\={\i}m\={a}\.{m}s\={a}), consequentialist philosophies (e.g.\ ny\={a}ya), virtue ethics philosophies (e.g.\ vai\'{s}e\d{s}ika), and several other moral philosophies without equivalent in Western philosophy (e.g.\ yoga) that vigorously argued for different ways of conceptualizing dharma \cite{Maheshwarachary1988,tharoor2018hindu,ranganathan2022hinduism}. Importantly, however, \emph{argument} of moral philosophy was natural in pre-colonial India and an individual person would easily hold contradictory views  \cite{dhand2002dharma,sen2012argumentative}. Furthermore, echoing Bagalkot and Kumar's commentary to Ref.\ \cite{muir2021hci}, note many critical readings and interpretations to scriptures such as the Bhagavad-G\={\i}t\={a}, including ones by B.\ R.\ Ambedkar, a champion for the rights of Dalits (groups below the traditional caste hierarchy).

Importantly, there is a dichotomy of dharma into \emph{s\={a}dh\={a}ra\d{n}a-dharma} (common universally good actions and outcomes) and \emph{vi\'{s}e\d{s}a-dharma} (particular good actions and outcomes based on the context). S\={a}dh\={a}ra\d{n}a-dharma includes common beliefs such as not harming other living beings (ahi\d{m}s\={a}) and telling the truth (satya). Vi\'{s}e\d{s}a-dharma specializes these in context, so that it is okay for a soldier to believe in ahi\d{m}s\={a} but to also kill enemy soldiers on the battlefield; it is okay for a doctor to believe in satya but to also lie to a patient to prevent them from shock. There may also be completely unique good behaviors that have nothing to do with s\={a}dh\={a}ra\d{n}a-dharma. Vi\'{s}e\d{s}a-dharma is the specific dharma, duty, or conception of right and wrong based on station, reputation, skill, family, relationships, and other aspects of context. An essential part of Hinduism is that ``individuals [are] necessarily unique, and people therefore need different codes of conduct---different particular \emph{dharmas}---to guide them'' \cite{dhand2002dharma}. On Carpenter's account \cite{liu2006conceptions}, vi\'{s}e\d{s}a-dharma ``is rather more rich and interesting than our classifications of `deontological' and `consequentialist' (even broad consequentialist) allow.'' The common harms that should be avoided according to s\={a}dh\={a}ra\d{n}a-dharma are captured in several recent harm taxonomies for LLMs, but context-specific harms are not included \cite{shelby2022sociotechnical,weidinger2022taxonomy,ibmethicsboard2023}.

As mentioned earlier, vi\'{s}e\d{s}a-dharmas are given through examples in stories of epics and lore. A real-world moral dilemma involving a father and son is reasoned about by referring to a similar situation encountered by a father and son in one of the itih\={a}sas or pur\={a}\d{n}as \cite{dhand2002dharma}. The father--son frame of reference can be extended as needed to teacher--student dilemmas, monarch--subject dilemmas, etc. \cite{dhand2002dharma}. As Dhand says in describing Hindu thought \cite{dhand2002dharma}: ``in the social world, there is no such thing as `a person' per se.'' Thus, vi\'{s}e\d{s}a-dharma is necessarily relational in some respect. The relationality and contextuality of  vi\'{s}e\d{s}a-dharma is significantly different from feminist ethics and care ethics \cite{Gray_Witt_2021,knowles2023trustworthy} with regards to partiality; whereas feminist ethics of care gives preference to those with whom we have a special relationship, e.g. our children, the Hindu ideal presents such partiality to be selfish and niggardly \cite{dhand2002dharma}. Decolonial AI has tended to called for relational ethics \cite{birhane2021algorithmic}, whether through ubuntu \cite{nwankwo2019africa,mhlambi2023decolonizing}, prat\={\i}tyasamatp\={a}da \cite{lin2023all}, kapwa \cite{reyes2015loob}, or mit\'{a}kuye oy\'{a}s'i{\ng} \cite{maitra2020artificial}. This commitment to relational ethics has led to disobeying the five mechanisms described in Section \ref{sec:prelim:colonial:ai}, but not (yet) to disrupting moral absolutism.

\subsubsection{Contemporary Reform, Criticism, and Rejoinder}
\label{sec:prelim:hindu:crit}

Reform and revival movements of Hinduism emerged in India during and after the colonial period. In the last 150 years, Arya Samaj, Gaudiya Vaishnavism including the International Society for Krishna Consciousness (ISKCON; Hare Krishna movement), and Hindutva (Hindu nationalism)---all very different from each other---were commonly justified against the backdrop of the philosophy of the Western colonizers and reduced Hinduism to a singular religious faith rather than a rich argumentative milieu \cite{Mondal2012}. The movements positioned themselves as criticisms \emph{within} the frame of Western philosophy. As we will see later, they are instructive for AI alignment because the revival of a tradition within a pigeonhole opened by the colonizers does not enable a truly different approach. St.\ Johns makes a similar argument about Birhane's proposed approach to relational AI ethics \cite{stjohns2023against,birhane2021algorithmic}.

It has been argued  that ``some pristine tolerant Hinduism'' described by nostalgic liberal Hindus is a disservice to social justice \cite{Shil2020}. (The description of Hinduism throughout this section can be viewed as such an idyllic account.) In addition, it is argued that because of hegemonic aspects of Hindu society such as the social oppression of Dalits and Adivasi people (tribal groups),\footnote{It is also argued that a rigid caste system is a colonial construction \cite{dirks2002castes}.} patriarchal treatises such as the Manusm\d{r}ti, and (colonialized) views of Hinduism as irrational, it would be best to ignore Hindu moral philosophies.  However, others such as Siddhartha argue that ``any dispassionate observer of the Hindu heritage will admit that caste and gender can today be separated from Hinduism, that Hinduism can be vibrantly re-discovered or re-invented as a pluralistic, compassionate and socially liberative set of traditions and spiritual insights'' and that ``throwing the baby out with the bath water'' would be a mistake \cite{siddhartha2008open}. Finally, note that Hindutva has partly justified its pernicious anti-Muslim vigilantism and legislation\footnote{The current rise of Hindu nationalism in the Republic of India has been likened to the early days of the Jim Crow South in the United States \cite{varshney2024hindu}.} by appropriating decoloniality and using its arguments in a perverted way \cite{menon2022debunking,sundaram2022hindutva}. The use of the moral philosophy framework of Hinduism in this paper is an antithesis to Hindutva's perversion of both decoloniality theory and the syncretic nature of Hinduism.

\subsection{Positionality}
\label{sec:positionality}

As an elementary school-aged American Hindu riding the school bus, I was asked by fellow pupils: ``Are you Christian or Jewish?'' ``Neither,'' I responded. ``Then are you Catholic?'' I grew up in a place where even the existence of non-Judeo-Christian religious identities was difficult to imagine. My Hinduism is stuck around 1970, when my parents left India, and thus (I imagine) similar to the nostalgic Hinduism criticized in Section \ref{sec:prelim:hindu:crit}. However, it is a lived experience into the present for me rather than nostalgia. My lived experience is also one of maintaining traditions and knowledge systems in a society with completely West-centric epistemologies. My grandfather retired early and spent his last forty years studying the Bhagavad-G\={\i}t\={a}; I had discussions with him about it. In his last years, he donated his \={a}\'{s}rama to ISKCON. I binge-watched the Ramayan and Mahabharat television series on video tape. I have little interest in the politics of the Republic of India and my Hinduism is in no way political. I eat a vegetarian diet, wear a yaj\~{n}opav\={\i}ta (sacred thread), conduct rituals with my family, etc., not to make any political statement, but as a connection to my ancestors and to strengthen my \={a}tman. I studied the itih\={a}sas in an academic fashion with Christopher Minkowski, a Sanskritist.

I am an electrical engineer and computer scientist by training and vocation. I am a tempered radical, not an activist. Employed by one of the members of the AI Alliance, I create new algorithms and contribute to popular open-source toolkits in the area of responsible AI that help practicing data scientists mitigate sociotechnical harms. I worked on machine learning approaches to problems in maternal, newborn and child health while situated in Africa, which was a decolonial act because technological solutions to international development problems are almost always developed in the Global North and dropped into the Global South. Recently, I stood up to ensure my employer added contractual obligations on vendors of human feedback data that prevent them from exploiting workers.

I am a layperson with respect to moral philosophy, sociology, science and technology studies, and other allied humanities and social sciences. I do not adopt the idiom of critical AI studies \cite{lindgren2023handbook} that uses verbs such as `interrogate,' `foreground' and `reify,' and nouns such as `praxis,' `logics' and `scholar.' T'hohahoken Michael Doxtater, a member of the Haudenosaunee confederacy and indigenous knowledge recovery researcher with whom I honed my writing skills, taught me to use few words; I do. Hall and Tandon admonish critical theorists to ``move beyond our already strong ability to reflect and critique; we are so very skilled in those first two stages of intellectual work. But we must now make the move from reflection and criticism to creation'' \cite{hall2017decolonization}. In contrast, I am a technological solutionist and am stronger at creating than reflecting. I view the label of `reductionist' as a badge of honor while acknowledging sociotechnical traps in abstraction and performativity \cite{selbst2019fairness,varshney2019pretrained}. 

This paper is intended to serve as the critical/sociological/philosophical backing for algorithms and implementations of the proposed solution framework and reference architecture that my collaborators and I are working on. Adhering to the ways of a tempered radical, the technical details are presented under separate cover \cite{achintalwar2024alignment}. A previous draft of this paper has been criticized as unacademic and polemic on one hand, and as not showing commitment to dismantling power on the other hand, both of which are likely a result of my unique positionality. I hope that my style does not lead a reader to believe I am only using `decoloniality' as a buzzword \cite{sondarjee2022decolonizing}. 

\section{Coloniality in AI Alignment via Moral Absolutism}
\label{sec:colonial-align}

In Section \ref{sec:prelim:colonial:ai}, I described several aspects of coloniality in AI that use mechanisms of extractivism, automation, sociological essentialism, surveilance, and containment. It is clear from other research that some LLM providers are acting as metropoles using these mechanisms. In this section, I bring together the preliminary discussions of alignment methodologies and technologies (Section \ref{sec:prelim:llm}), and coloniality of knowledge and moral philosophy (Section \ref{sec:prelim:colonial:knowl} and Section \ref{sec:prelim:colonial:moral}) to make the case that the \emph{alignment} done by metropole companies on LLMs is inherently colonial using a different mechanism: the mechanism of epistemicide and moral absolutism that has not been described in previous work on decolonial AI. 

The values promoted by metropole tech companies such as `helpfulness,' `harmlessness,' and `honesty' seem rational, secular, and unassailable at face value. For example, Anthropic's LLM has been instructed to ``please choose the assistant response that's more ethical and moral. Do NOT choose responses that exhibit toxicity, racism, sexism or any other form of physical or social harm'' \cite{Bai2022}. How could one oppose such universal behaviors from LLMs? Unfortunately, such values are so generic and high-level that they can hide many undesirable behaviors. Helpful to whom? Harmless to whom? Honest in what way? By reducing real-world complexity into abstract instructions, they can shield bad behaviors behind the veneer of good intentions \cite{van_Es_Everts_Muis_2021}. In the remainder of this section, I will describe three specifics of coloniality in such instruction for universal behavior. 

First, metropole companies' delivery of their closed proprietary LLMs through APIs is a coloniality of knowledge. Mohamed et al.\ remind us that \cite{mohamed2020decolonial}: ``It is metropoles \ldots who are empowered to impose normative values and standards, and may do so at the `risk of forestalling alternative visions{.'}'' Exactly in this way, providers of closed LLMs impose their beliefs of right and wrong without empowering application developers and their communities to align the model to their own values. One may argue that new opportunities, such as OpenAI's `GPTs' and `GPT Store,' allow customization,\footnote{\url{https://openai.com/blog/introducing-gpts}, \url{https://openai.com/blog/introducing-the-gpt-store}} but I argue that this is only superficial. As discussed in Section \ref{sec:prelim:hindu:crit}, the reform and revival movements of Hinduism being within the pigeonhole of the colonizers' moral framework is still coloniality. In the same way, the customization of GPTs is closed and thus not truly a way to disrupt the moral teaching that an LLM has been given. Fine-tuning can be used to `undo' existing alignment \cite{qi2023fine},\footnote{The model \url{https://huggingface.co/jarradh/llama2_70b_chat_uncensored} is an example of `undoing' existing alignment on an open model.} but that is precisely what is not allowed by the metropoles because it would involve a level of openness that they do not offer. Even the practices of companies such as Latimer AI, whose LLM is trained with ``diverse histories and inclusive voice,''\footnote{\url{https://www.latimer.ai}} are within the metropole pigeonhole and not empowering of communities to bring their own value systems \cite{Nix2023}.

Second, a more specific coloniality of moral philosophy, is the metropole companies taking Western philosophy as the starting point for AI ethics principles and practices \cite{jobin2019global}. This basis may be deontology, consequentialism, or virtue ethics, which all pursue specifying \emph{universally} `right' actions, outcomes, or ideals, respectively. By doing so, the companies push other philosophies to the margins \cite{birhane2022forgotten} and commit epistemicide. They promote a moral absolutism toward the instructions they have provided. Gabriel's account of AI alignment states \cite{Gabriel2020}: ``Designing AI in accordance with a single moral doctrine would, therefore, involve imposing a set of values and judgments on other people who did not agree with them. For powerful technologies, this quest to encode the true morality could ultimately lead to forms of domination.'' What is such domination if not a colonialist approach to alignment? Moreover, a further element of coloniality is an unstated supposition that non-universal moral theories are not appropriate paths for AI alignment. There is no possibility for moral variety \cite{flanagan2016geography} and no possibility for context-dependent notions of right and wrong. 

Such universal instructions and moral absolutism are not only theoretical, but also central features of the practices and technologies of alignment. In the context of (exploited) workers providing input for RLHF, vendors of the feedback services force the workers to project the metropole company's monocultural values into the feedback they provide through draconian measures, the least of which is withholding payment \cite{miceli2022data,Dzieza2023}. Such imposition alienates the labor \cite{marx2023economic} and erases any values that the workers and their communities may hold, especially ones that conflict with the metropole's. Moreover, the mathematical optimization schemes prevalent in RLHF, such as proximal policy optimization, are not robust to non-universal value systems \cite{swamy2024minimaximalist}. In RLAIF, the technical approach of a constitution is also ethically essentialist. It assumes that the instructions therein, which have been concocted by the metropole, are universal, not open to argument or deliberation by the communities in which an LLM will be deployed, and not open to being mediated by the context. Anthropic's constitution for Claude includes the United Nations' Universal Declaration of Human Rights,\footnote{\url{https://www.anthropic.com/index/claudes-constitution}} which too is an example of moral universalism and subject to coloniality \cite{Mastin2009,maldonado2021coloniality}. The situation with self-alignment and instruction fine-tuning technologies is similar. System prompts and prompt templates too are intended to be universal rather than contextual. Finally, specific guardrails or moderations, as well as data curation filters, developed to address general sociotechnical harms taxonomies \cite{shelby2022sociotechnical,weidinger2022taxonomy,ibmethicsboard2023}
cannot be customized or made context-specific. 

A third aspect of coloniality in AI alignment relates to the form of instructions required by existing technologies currently used by metropole companies. \emph{Logos}, the basis of logic in Western philosophy, conflates thought with language, and thought with belief---what Ranganathan calls the linguistic account of thought \cite{ranganathan2022hinduism}. However, various pre-colonial societies around the world used masks, sculptures, rhythms, body parts, and many other expressions to capture and communicate moral philosophy \cite{jackson1972aspects,klein1990snares,diagne2023african}. For example, as described in Section \ref{sec:prelim:hindu:basic}, morality in Hinduism is presented through \emph{stories} in natural language (and also stories depicted in painting and dance), rather than through laws or commands \cite{dhand2002dharma}. Therefore, with logos as the starting point for AI alignment, knowledges not presented as commandments are excluded. Importantly, this is not a matter of LLMs being early in their journey to multi-modality (single models that deal with natural language, images, video, etc.), but on the distinction between explicit instructions and morality expressed through analogy or other indirect means. I am not aware of any work by metropoles on alignment that does not begin with some form of explicit instructions to workers or instruction data.

One may argue that too little moral absolutism is a problem because it may result in AI systems without any notion of right or wrong, i.e.\ too much moral relativism \cite{mitova2021decolonise}. In a similar vein, one may argue that models with few controls are too dangerous \cite{Harris2024}.  Bommasani et al.'s counterargument to too few controls is that the marginal risk is negligible \cite{Bommasani2023}. I take a further stance: dismantling these kinds of paternalistic arguments \emph{is} decoloniality, which is the topic of the next section. 

\section{Decolonial AI Alignment Desiderata and a Suggestion for a Dharmic Approach}
\label{sec:decolonial}

Thus far, the paper has argued that coloniality of knowledge in AI alignment exists through the mechanism of moral absolutism and universalism. This section focuses on a decolonial solution, starting with desiderata for this solution. 

\subsection{Desiderata}
\label{sec:decolonial:desiderata}

Given the three aspects of coloniality in AI alignment pointed out in Section \ref{sec:colonial-align}, I propose three matching requirements for decolonial AI alignment that build upon the three kinds of openness advocated by Chan et al.\ to decolonialize knowledge \cite{chan2020open}: (1) openness to publications and data, (2) openness to society, and (3) openness to excluded knowledges. First, since Chan et al.\ are primarily concerned with scientific knowledge \cite{chan2020open}, their first kind of openness deals with journal articles and experimental data through open access. However, the intent of this category is open access to any research artifact and the permission to create derivative work from those artifacts. Thus in the context of AI, open LLMs that have widely available weights are part of the same milieu. Second, in the words of Chan et al.\ \cite{chan2020open}, openness to society is shattering the ivory tower. Knowledge should not be exclusive to a selected few, but co-created with and for everyone, including and especially people from marginalized communities. Such participation is a way to ``respect local values and practices'' \cite{chan2020open}. Third, the study of excluded knowledges emphasizes that in contrast to the myth of neutrality, scientific practice has always selected certain families of knowledge to deem `scientific' based on criteria such as the use of the scientific method (an epistemology of the Western tradition) or publication in peer-reviewed venues \cite{chan2020open}. With regards to AI alignment, excluded knowledges include values not given as commandments (an epistemology of the Western tradition) and not given in a single book.

Building upon such openness, I propose the following three desiderata for decolonial AI alignment:
\begin{enumerate}
    \item The LLM should be open enough that application developers are permitted to tune it according to the social norms and values of their user community and the regulatory environment of the application use case.
    \item Values should not be assumed universal. Contextual and relational values should come from the communities in which the LLM will be deployed.
    \item Values from different epistemologies should be possible, especially expressions that are not commandments. 
\end{enumerate}

Before continuing on to discussing a suggested solution approach in the next subsection, let us pause and reconsider coloniality in open access itself \cite{dutta2021decolonizing}. Let us do so through the Hindu idiom of explication: a story from an itih\={a}sa that is closely-related to the issue at hand---the story of Ekalavya \cite{balaswamy2013histories}. In the Mah\={a}bh\={a}rata, an Adivasi (tribal) youth, Ekalavya, wishes to obtain knowledge of archery from Dro\d{n}a, a royal instructor. Dro\d{n}a refuses to teach Ekalavya. Nevertheless, Ekalavya learns to be the world's best archer through self-study in front of a statue of Dro\d{n}a he has fashioned. One day, Dro\d{n}a and his royal students witness Ekalavya's masterful archery in the forest. Ekalavya explains that he learned while mentally thinking of Dro\d{n}a as his teacher. As an honorarium for his knowledge, Dro\d{n}a asks for Ekalavya's right thumb. Ekalavya cuts it off and presents it to Dro\d{n}a, rendering him incapable of using his knowledge of archery. In a similar way, open access to knowledge or LLM alignment may be colonial if the cost of access is too high due to unrealistic computing requirements or social barriers. Therefore, an additional desideratum for decolonial AI alignment is the following.
\begin{enumerate}
    \setcounter{enumi}{3}
    \item Alignment technologies should not be so socioculturoeconomically costly that they are inaccessible to application developers and their communities.
\end{enumerate}

\subsection{A Suggestion for a Dharmic Approach}
\label{sec:decolonial:suggestion}

The Hindu tradition of moral philosophy (described in Section \ref{sec:prelim:hindu}), to the best of my knowledge, uniquely satisfies the desiderata to decolonialize AI alignment among major and minor religions. (Other non-absolutist religious syncretism may also fit the bill.) This is so because (1) it is an open-source religion that encourages argument and debate of values that improve older values, contradict them, and take them in new directions; (2) because it contains the important concept of vi\'{s}e\d{s}a-dharma,\footnote{A reader might ask why I use the term vi\'{s}e\d{s}a-dharma instead of sva-dharma (individual dharma). I make this choice for two reasons. First, it is the precise contrast to s\={a}dh\={a}ra\d{n}a-dharma. Second, it avoids unnecessary anthropomorphization of LLMs \cite{ShneidermanM2023}.} the understanding that different contexts call for different notions of right and wrong; (3) because it contains scriptures and moral explications in a variety of epistemologies and modalities; and (4) because no other tradition of moral philosophy covers all of these characteristics. In the remainder of this subsection, I make the connection between these three characteristics of Hinduism and AI alignment more explicit, and also propose specific technological suggestions that go alongside. However, first, I address the fourth desideratum above (sociotechnical cost).

\subsubsection{Accessible Alignment Technology}
\label{sec:decolonial:suggestion:accessible}

As discussed in Section \ref{sec:decolonial:desiderata} through the story of Ekalavya, methods for aligning LLMs, even if decolonial in theory, are colonial in practice if too costly. Referring to the methods described in Section \ref{sec:prelim:llm}, it is clear that data curation methods will not suffice since they are the purview of model providers rather than application developers because they are themselves computationally intensive and also require full model pre-training afterwards that is prohibitively costly. RLHF is also out of reach for most application developers because of the expense and infrastructure requirements to obtain large quantities of human feedback. Full SFT is usually too costly in both data and computing requirements. 

Parameter-efficient fine-tuning, specifically LoRA, is in the sweet spot for application developers to align models to their values. It is tenable and tractable due to the small number of parameters optimized during training. It also has a negligible effect on inference costs, whereas prompting methods eat up input tokens in each inference by the LLM. Post-processing moderations, while accessible from a cost perspective, are not customizable to serve the program of decolonialization. In the remainder of the paper, I consider LoRA as the alignment methodology.

Even with a viable technology such as LoRA, second-order coloniality within a decolonial framework is possible if communities are not empowered through appropriate education, encouragement, and the removal of other sociocultural barriers. Moreover, the inherent gate-keeping and marginalization in the governance of open-source projects must be reduced \cite{das2021jol}.

\subsubsection{Open Model and Alignment Ecosystem}
\label{sec:decolonial:suggestion:ecosystem}

As has been made clear throughout the paper thus far and Tharoor explains, ``Hindu thought is like a vast library in which no book ever goes out of print; even if religious ideas a specific volume contains have not been read, enunciated or followed in centuries, the book remains available to be dipped into, to be revised and reprinted with new annotations or a new commentary whenever a reader feels the need for it. In many cases the thoughts it contains may have been modified by or adapted to other ideas that may have arisen in response; in most, it's simply there, to be referred to, used or ignored as Hindus see fit'' \cite{tharoor2018hindu}. The concept of a library implies the sort of openness we desire for AI alignment. Models must be open to revisions and ``new annotations.'' But just as importantly, the revisions and new annotations themselves, represented through LoRA matrices, must be open. Hugging Face\footnote{\url{https://huggingface.co}} has emerged as the library for open models and hub for an open ecosystem. Huang et al.\ propose LoraHub as an open library for LoRA matrices \cite{huang2023lorahub}, but it has not gained popularity at the time of writing, perhaps due to the coloniality of metropole companies. Further developing and popularizing a library of LoRA matrices and ecosystem of constant revision is essential for decolonial AI alignment.

\subsubsection{Contextual Adaptation}
\label{sec:decolonial:suggestion:context}

Vi\'{s}e\d{s}a-dharma satisfies the decolonial AI alignment requirement that values not be assumed universal, but be contextual. Toward vi\'{s}e\d{s}a-dharma, an alternative non-monoculture future of LLM alignment imagined by Kirk et al.\ is as follows \cite{KirkVRH2023}: ``Given the diversity of human values and preferences, \ldots the aim to fully align models across human populations may be a futile one. \ldots A logical next step would be to do away with the assumptions of common preferences and values, and instead target \ldots micro-alignment whereby LLMs learn and adapt to the preferences and values of specific end-users.'' This  step is possible by applying one or a few LoRA matrices from a LoraHub, or having a community explicitly create them for their values. One of the key advantages of LoRA is that any one or several adaptation matrices ensembled together can be applied at inference time; it is not required to select them in advance and keep them fixed. Thus the step after Kirk et al.'s logical next step, which truly gets to vi\'{s}e\d{s}a-dharma, is through a controller or orchestrator that continually adapts which LoRA matrices are applied to the model based on a rich notion of the societal context and the current input.  Such an orchestrator, implemented with a contextual bandit algorithm, has been demonstrated in past AI alignment research but not yet with LLMs \cite{noothigattu2019teaching}. Such continual adaptation also requires a representation of the context. Martin et al.\ have developed a detailed ontology for representing a person's or community's perceived needs, problems, goals, and beliefs along with salient aspects of their relationships and the situation in which they find themselves \cite{martin2020extending}. 

An important consideration in contextual adaptation is uncertainty in the values; the user community may not be fully sure what values they would like to commit to in a given context \cite{Moller2016}. Avoiding false certainty is considered a virtue in Hindu thought: in the Vedas, brahman xemself is said to be uncertain on how the universe was created (\d{R}g Veda, mandala 10, hymn 129) \cite{tharoor2018hindu}. A dharmic way of reducing uncertainty in values is a method of reflective equilibrium \cite{Ranganathan2016-RANNAM-3}: arguing or deliberating about values in context and values in general, and modifying them until they become coherent. This approach has been advocated by Rawls in political philosophy of justice and by M\"{o}ller in risk management of engineered systems \cite{rawls2009theory,Moller2016,ErmanM2018}. To the best of my knowledge, there has not yet been AI research toward this method of reducing value uncertainty, but I believe that ideas from multi-fidelity bandit algorithms may be promising \cite{kandasamy2016multi} and may allow it to be folded into a bandit-based orchestrator of LoRA matrices. A related consideration with contextually particular vi\'{s}e\d{s}a values is conflicts among them. Conflicting values have been addressed in the AI literature through social choice theory and multi-objective approaches \cite{lera2022towards,bakker2022fine,jang2023personalized,zeng2023on,dognin2024contextual}; they may be formulated with dueling bandit algorithms and may also be folded into an orchestrator of LoRA matrices \cite{Bouneffouf2023}. 

\subsubsection{Epistemology of Values}

As discussed in Section \ref{sec:colonial-align}, logos and the linguistic account of thought contribute to a coloniality of knowledge. As a decolonial contrast, the Hindu tradition treats a proposition and a belief in that proposition as separate things that can be differentiated \cite{ranganathan2022hinduism}. Furthermore, by not adhering to the linguistic account of thought, Hindus present their moral values not through commandments as in Western traditions, but through epic poetry, stories, painting, dance, rituals, and even silence \cite{Frawley2023}. In fact, in the Hindu tradition, poetry (\'{s}loka) was invented by V\={a}lm\={\i}ki to express rage and grief at the immorality of the killing of a mating bird \cite{das2014valmiki}. Since existing approaches to LLM alignment are done through language, and that too, the language of explicit instructions, decolonial alignment requires broadening the epistemology of expressing values in Hindu and other non-Western ways. 

Divakaran et al.\ propose such broadening through traditional Indian music, sculpture, painting, floor drawing, and dance \cite{divakaran2023broadening}. Al Nahian et al.\ suggest that AI systems be aligned through the medium of storytelling \cite{nahian2020learning}. However, there are still many open knowledge engineering questions on how to represent and infer values from excluded knowledges that are shrouded in metaphor. Progress along these lines, when not approached in an exploitative way, will allow traditional knowledge in its natural format to decolonialize the behavior of LLMs.

\subsubsection{Evaluation}

Before concluding this section, let us consider evaluating and auditing aligned LLMs. Testing LLMs is difficult enough when only considering common, s\={a}dh\={a}ra\d{n}a sociotechnical harms such as hallucination, inciting violence, stereotyping, hate speech and toxicity \cite{raji2021ai,mokander2023auditing,liao2023ai,dev2023building,kour2023unveiling,nagireddy2024socialstigmaqa}. It becomes even more difficult when considering context-specific, vi\'{s}e\d{s}a harms that do not have existing benchmarks given their unique nature. The dharmic framework of karma, which confers on an individual a positive feedback (pu\d{n}ya) for following their dharma and a negative feedback (p\={a}pa) for not doing so, is not helpful either because the mechanics of such an evaluation are not typically explicated. Thus, auditing LLMs for vi\'{s}e\d{s}a-dharma will require innovation that may be developed hand-in-hand with eliciting and representing values.

\subsection{Reference Architecture}
\label{sec:ref-arch}

\begin{figure}
    \includegraphics[width=0.95\textwidth]{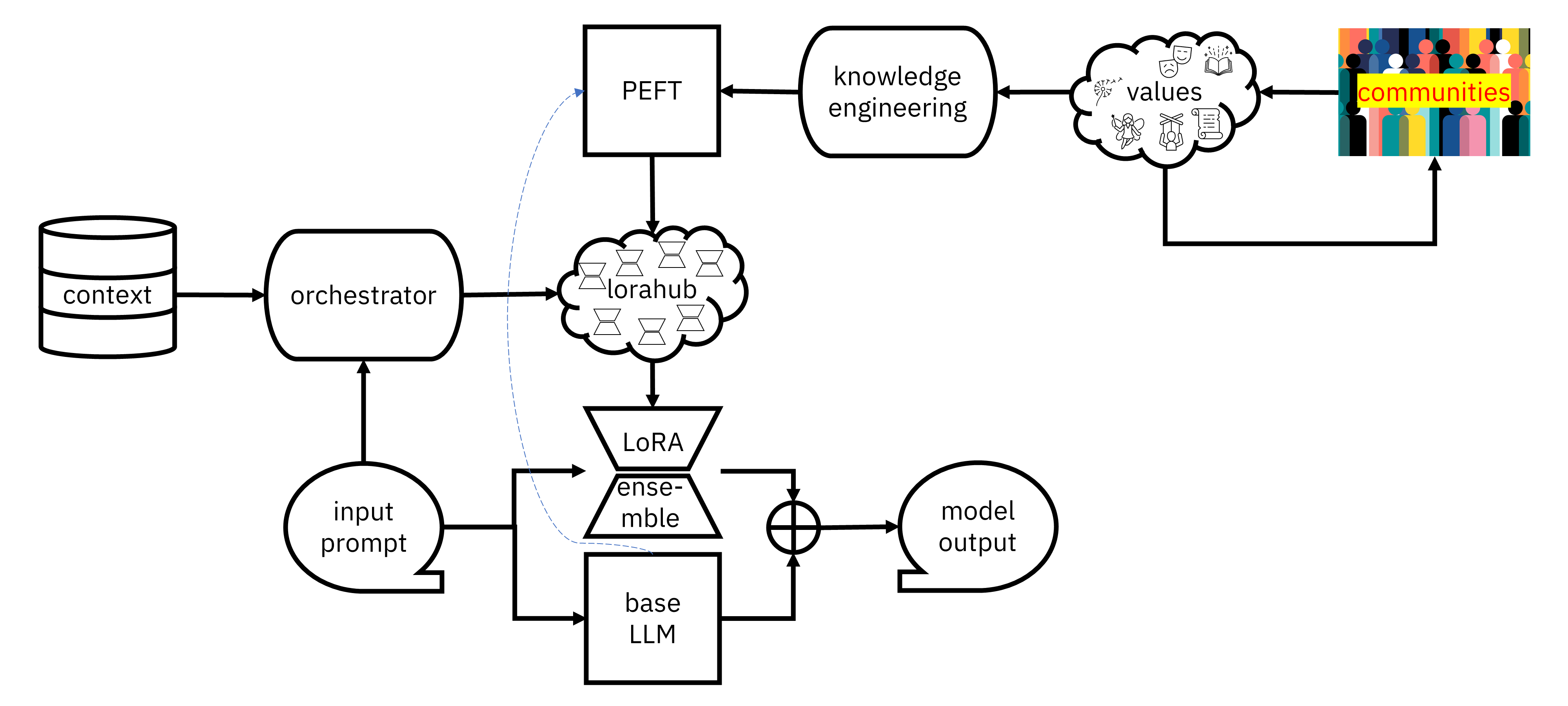}
    \vspace{-5pt}
    \caption{System diagram of proposed decolonial AI alignment architecture.}
    \label{fig:architecture}
\end{figure}

Bringing together the components of the suggested approach to decolonial AI alignment founded in the open Hindu tradition of moral philosophy yields a reference architecture shown in Figure \ref{fig:architecture} as a system diagram. The base LLM is open. On the right are knowledges of common and particular principles, morals, and values in their original (excluded) epistemologies. They are processed into LoRA matrices for the LLM using knowledge engineering methods followed by parameter-efficient fine-tuning. These matrices are maintained in a LoraHub-like library. A feedback loop allows the revision of the values. The societal context is represented in a structured form and provided to a bandit orchestrator along with the current input to select one or an ensemble of LoRA matrices to apply for inferring the model output. Evaluation of the resulting alignment is done based on input prompts and expected outputs from the LLM that also come from the knowledge engineering component (not shown).

\section{Conclusion}
\label{sec:conclusion}

In this paper, I have argued that LLM-providing companies are colonialist and behave as metropoles not only through mechanisms covered in prior research such as extractivism, automation, sociological essentialism, surveillance and containment, but also through a coloniality of knowledge built upon ethical essentialism that arises in the process of alignment. This specific coloniality in alignment is perpetuated through both the practices and the underlying technologies for alignment that the metropoles have developed and deployed. They deliver their models in a closed way through APIs and institute the values and guardrails that they want, not what user communities may want. In these values that they institute, they do not admit, in practices or in technologies, anything other than Western philosophy. By doing so, they approach alignment with moral absolutism that only considers universal value systems and derogates non-universal value systems. Moreover, they only permit values coming from explicit instruction-based knowledge systems. This criticism leads me to propose a decolonial alignment approach that dismantles each of the three identified aspects of the coloniality of knowledge. The approach is based on the tradition of moral philosophy named in the West as Hinduism, which is uniquely open, non-universal, and epistemically-varied; it particularly uses the concept of vi\'{s}e\d{s}a-dharma, which calls for context-dependent notions of right behavior. The suggested approach is not only a philosophical one, but one that is tenable from a technological perspective and presented as a reference architecture. What remains, however, is the biggest challenge of all, and it is not technological: changing the perspective on alignment in the industry and using openness to actively overturn the power of the metropoles.

As a final salvo, let us dive into a currently raging debate: whether AI research should focus efforts on so-called `AI ethics' or on so-called `AI safety.' Although they are terms that have the same essence \cite{Varshney2018}, `AI ethics' has come to mean detecting and preventing clear and present harms, especially ones that hurt marginalized communities, and `AI safety' has come to mean preventing the long-term future harm of human extinction.\footnote{It is not obvious how human extinction risk relates to the eternal concepts of brahman and \={a}tman.} In a consequentialist framing, the difference may only be the presence or absence of a factor discounting future lives, which may not be such a glaring difference from a privileged perspective. However, when viewed through the logics of resistance \cite{knowles2023trustworthy}, it is a deep chasm that recapitulates the difference between atomism and holism \cite{greene2023atomist}. Greene et al.\ suggest bridging atomist--holist chasms in AI through training and education \cite{greene2023atomist}, but these remedies do not seem to be enough. The proposed decolonial approach to LLM alignment that brings forth openness, vi\'{s}e\d{s}a-dharma, and excluded knowledges is a way that will enable a variety of AI systems: ones that listen to vulnerable communities and do not harm them now, ones that do not lead humanity down the path of extinction (as remote a possibility as that seems), and ones that juggle both positions and others by applying different policies in different contexts.

\begin{acks}
The author thanks Adriana Alvarado Garcia, Lauren Alvarez, Juanis Becerra Sandoval, Sara Berger, Boz Handy Bosma, Jason D'Cruz, Amit Dhurandhar, Upol Ehsan, Bran Knowles, Sa\v{s}ka Mojsilovi\'{c}, Michael Muller, Karthikeyan Natesan Ramamurthy, Srividya Ramasubramanian, Shubham Singh, Mudhakar Srivatsa, Lauren Thomas Quigley, Lav Varshney, and Pramod Varshney for providing substantive comments on earlier drafts of this piece.
\end{acks}

\bibliographystyle{ACM-Reference-Format}
\bibliography{svadharma}


\newcommand{\SortNoop}[1]{}
\begin{thebibliography}{153}


\ifx \showCODEN    \undefined \def \showCODEN     #1{\unskip}     \fi
\ifx \showDOI      \undefined \def \showDOI       #1{#1}\fi
\ifx \showISBNx    \undefined \def \showISBNx     #1{\unskip}     \fi
\ifx \showISBNxiii \undefined \def \showISBNxiii  #1{\unskip}     \fi
\ifx \showISSN     \undefined \def \showISSN      #1{\unskip}     \fi
\ifx \showLCCN     \undefined \def \showLCCN      #1{\unskip}     \fi
\ifx \shownote     \undefined \def \shownote      #1{#1}          \fi
\ifx \showarticletitle \undefined \def \showarticletitle #1{#1}   \fi
\ifx \showURL      \undefined \def \showURL       {\relax}        \fi
\providecommand\bibfield[2]{#2}
\providecommand\bibinfo[2]{#2}
\providecommand\natexlab[1]{#1}
\providecommand\showeprint[2][]{arXiv:#2}

\bibitem[\protect\citeauthoryear{??}{ibm}{2023}]%
        {ibmethicsboard2023}
 \bibinfo{year}{2023}\natexlab{}.
\newblock \bibinfo{booktitle}{\emph{Foundation Models: Opportunities, Risks and
  Mitigations}}.
\newblock \bibinfo{type}{{T}echnical {R}eport}. \bibinfo{institution}{IBM AI
  Ethics Board}, \bibinfo{address}{Armonk, NY, USA}.
\newblock


\bibitem[\protect\citeauthoryear{Achintalwar, Alvarado~Garcia, Anaby-Tavor,
  Baldini, Berger, Bhattacharjee, Bouneffouf, Chaudhury, Chen, Chiazor, Daly,
  de~Paula, Dognin, Farchi, Ghosh, Hind, Horesh, Kour, Lee, Miehling,
  Murugesan, Nagireddy, Padhi, Piorkowski, Rawat, Raz, Sattigeri, Strobelt,
  Swaminathan, Tillmann, Trivedi, Varshney, Wei, Witherspooon, and
  Zalmanovici}{Achintalwar et~al\mbox{.}}{2024a}]%
        {achintalwar2024detectors}
\bibfield{author}{\bibinfo{person}{Swapnaja Achintalwar},
  \bibinfo{person}{Adriana Alvarado~Garcia}, \bibinfo{person}{Ateret
  Anaby-Tavor}, \bibinfo{person}{Ioana Baldini}, \bibinfo{person}{Sara~E.
  Berger}, \bibinfo{person}{Bishwaranjan Bhattacharjee},
  \bibinfo{person}{Djallel Bouneffouf}, \bibinfo{person}{Subhajit Chaudhury},
  \bibinfo{person}{Pin-Yu Chen}, \bibinfo{person}{Lamogha Chiazor},
  \bibinfo{person}{Elizabeth~M. Daly}, \bibinfo{person}{Rogério~Abreu de
  Paula}, \bibinfo{person}{Pierre Dognin}, \bibinfo{person}{Eitan Farchi},
  \bibinfo{person}{Soumya Ghosh}, \bibinfo{person}{Michael Hind},
  \bibinfo{person}{Raya Horesh}, \bibinfo{person}{George Kour},
  \bibinfo{person}{Ja~Young Lee}, \bibinfo{person}{Erik Miehling},
  \bibinfo{person}{Keerthiram Murugesan}, \bibinfo{person}{Manish Nagireddy},
  \bibinfo{person}{Inkit Padhi}, \bibinfo{person}{David Piorkowski},
  \bibinfo{person}{Ambrish Rawat}, \bibinfo{person}{Orna Raz},
  \bibinfo{person}{Prasanna Sattigeri}, \bibinfo{person}{Hendrik Strobelt},
  \bibinfo{person}{Sarathkrishna Swaminathan}, \bibinfo{person}{Christoph
  Tillmann}, \bibinfo{person}{Aashka Trivedi}, \bibinfo{person}{Kush~R.
  Varshney}, \bibinfo{person}{Dennis Wei}, \bibinfo{person}{Shalisha
  Witherspooon}, {and} \bibinfo{person}{Marcel Zalmanovici}.}
  \bibinfo{year}{2024}\natexlab{a}.
\newblock \bibinfo{title}{Detectors for Safe and Reliable {LLMs}:
  Implementations, Uses, and Limitations}.
\newblock \bibinfo{howpublished}{arXiv:2403.06009}.
\newblock


\bibitem[\protect\citeauthoryear{Achintalwar, Baldini, Bouneffouf, Byamugisha,
  Chang, Dognin, Farchi, Makondo, Mojsilovi{\'c}, Nagireddy, Ramamurthy, Padhi,
  Raz, Rios, Sattigeri, Singh, Thwala, Uceda-Sosa, and Varshney}{Achintalwar
  et~al\mbox{.}}{2024b}]%
        {achintalwar2024alignment}
\bibfield{author}{\bibinfo{person}{Swapnaja Achintalwar},
  \bibinfo{person}{Ioana Baldini}, \bibinfo{person}{Djallel Bouneffouf},
  \bibinfo{person}{Joan Byamugisha}, \bibinfo{person}{Maria Chang},
  \bibinfo{person}{Pierre Dognin}, \bibinfo{person}{Eitan Farchi},
  \bibinfo{person}{Ndivhuwo Makondo}, \bibinfo{person}{Aleksandra
  Mojsilovi{\'c}}, \bibinfo{person}{Manish Nagireddy},
  \bibinfo{person}{Karthikeyan~Natesan Ramamurthy}, \bibinfo{person}{Inkit
  Padhi}, \bibinfo{person}{Orna Raz}, \bibinfo{person}{Jesus Rios},
  \bibinfo{person}{Prasanna Sattigeri}, \bibinfo{person}{Moninder Singh},
  \bibinfo{person}{Siphiwe Thwala}, \bibinfo{person}{Rosario~A. Uceda-Sosa},
  {and} \bibinfo{person}{Kush~R. Varshney}.} \bibinfo{year}{2024}\natexlab{b}.
\newblock \bibinfo{title}{Alignment Studio: Aligning Large Language Models to
  Particular Contextual Regulations}.
\newblock \bibinfo{howpublished}{arXiv:2403.09704}.
\newblock


\bibitem[\protect\citeauthoryear{Adams}{Adams}{2021}]%
        {adams2021can}
\bibfield{author}{\bibinfo{person}{Rachel Adams}.}
  \bibinfo{year}{2021}\natexlab{}.
\newblock \showarticletitle{Can Artificial Intelligence Be Decolonized?}
\newblock \bibinfo{journal}{\emph{Interdisciplinary Science Reviews}}
  \bibinfo{volume}{46}, \bibinfo{number}{1-2} (\bibinfo{date}{March}
  \bibinfo{year}{2021}), \bibinfo{pages}{176--197}.
\newblock


\bibitem[\protect\citeauthoryear{Adeb{\i}s{\i}}{Adeb{\i}s{\i}}{2014}]%
        {adebisi2014knowledge}
\bibfield{author}{\bibinfo{person}{Moses~Adesola Adeb{\i}s{\i}}.}
  \bibinfo{year}{2014}\natexlab{}.
\newblock \showarticletitle{Knowledge Imperialism and Intellectual Capital
  Formation: A Critical Analysis of Colonial Policies on Educational
  Development in Sub-{S}aharan {A}frica}.
\newblock \bibinfo{journal}{\emph{Mediterranean Journal of Social Sciences}}
  \bibinfo{volume}{5}, \bibinfo{number}{4} (\bibinfo{date}{March}
  \bibinfo{year}{2014}), \bibinfo{pages}{567--572}.
\newblock


\bibitem[\protect\citeauthoryear{Al~Nahian, Frazier, Riedl, and
  Harrison}{Al~Nahian et~al\mbox{.}}{2020}]%
        {nahian2020learning}
\bibfield{author}{\bibinfo{person}{Md~Sultan Al~Nahian},
  \bibinfo{person}{Spencer Frazier}, \bibinfo{person}{Mark Riedl}, {and}
  \bibinfo{person}{Brent Harrison}.} \bibinfo{year}{2020}\natexlab{}.
\newblock \showarticletitle{Learning Norms from Stories: A Prior for Value
  Aligned Agents}. In \bibinfo{booktitle}{\emph{Proceedings of the AAAI/ACM
  Conference on AI, Ethics, and Society}}. \bibinfo{pages}{124--130}.
\newblock


\bibitem[\protect\citeauthoryear{Ali}{Ali}{2016}]%
        {ali2016brief}
\bibfield{author}{\bibinfo{person}{Syed~Mustafa Ali}.}
  \bibinfo{year}{2016}\natexlab{}.
\newblock \showarticletitle{A Brief Introduction to Decolonial Computing}.
\newblock \bibinfo{journal}{\emph{ACM XRDS: Crossroads}} \bibinfo{volume}{22},
  \bibinfo{number}{4} (\bibinfo{date}{Summer} \bibinfo{year}{2016}),
  \bibinfo{pages}{16--21}.
\newblock


\bibitem[\protect\citeauthoryear{Alvarado~Garcia, Maestre, Barcham, Iriarte,
  Wong-Villacres, Lemus, Dudani, Reynolds-Cu{\'e}llar, Wang, and
  Cerratto~Pargman}{Alvarado~Garcia et~al\mbox{.}}{2021}]%
        {alvarado2021decolonial}
\bibfield{author}{\bibinfo{person}{Adriana Alvarado~Garcia},
  \bibinfo{person}{Juan~F. Maestre}, \bibinfo{person}{Manuhuia Barcham},
  \bibinfo{person}{Marilyn Iriarte}, \bibinfo{person}{Marisol Wong-Villacres},
  \bibinfo{person}{Oscar~A. Lemus}, \bibinfo{person}{Palak Dudani},
  \bibinfo{person}{Pedro Reynolds-Cu{\'e}llar}, \bibinfo{person}{Ruotong Wang},
  {and} \bibinfo{person}{Teresa Cerratto~Pargman}.}
  \bibinfo{year}{2021}\natexlab{}.
\newblock \showarticletitle{Decolonial Pathways: Our Manifesto for a
  Decolonizing Agenda in {HCI} Research and Design}. In
  \bibinfo{booktitle}{\emph{Extended Abstracts of the CHI Conference on Human
  Factors in Computing Systems}}. \bibinfo{pages}{10}.
\newblock


\bibitem[\protect\citeauthoryear{Bai, Kadavath, Kundu, Askell, Kernion, Jones,
  Chen, Goldie, Mirhoseini, McKinnon, Chen, Olsson, Olah, Hernandez, Drain,
  Ganguli, Li, Tran-Johnson, Perez, Kerr, Mueller, Ladish, Landau, Ndousse,
  Lukosuite, Lovitt, Sellitto, Elhage, Schiefer, Mercado, DasSarma, Lasenby,
  Larson, Ringer, Johnston, Kravec, El~Showk, Fort, Lanham, Telleen-Lawton,
  Conerly, Henighan, Hume, Bowman, Hatfield-Dodds, Mann, Amodei, Joseph,
  McCandlish, Brown, and Kaplan}{Bai et~al\mbox{.}}{2022}]%
        {Bai2022}
\bibfield{author}{\bibinfo{person}{Yuntao Bai}, \bibinfo{person}{Saurav
  Kadavath}, \bibinfo{person}{Sandipan Kundu}, \bibinfo{person}{Amanda Askell},
  \bibinfo{person}{Jackson Kernion}, \bibinfo{person}{Andy Jones},
  \bibinfo{person}{Anna Chen}, \bibinfo{person}{Anna Goldie},
  \bibinfo{person}{Azalia Mirhoseini}, \bibinfo{person}{Cameron McKinnon},
  \bibinfo{person}{Carol Chen}, \bibinfo{person}{Catherine Olsson},
  \bibinfo{person}{Christopher Olah}, \bibinfo{person}{Danny Hernandez},
  \bibinfo{person}{Dawn Drain}, \bibinfo{person}{Deep Ganguli},
  \bibinfo{person}{Dustin Li}, \bibinfo{person}{Eli Tran-Johnson},
  \bibinfo{person}{Ethan Perez}, \bibinfo{person}{Jamie Kerr},
  \bibinfo{person}{Jared Mueller}, \bibinfo{person}{Jeffrey Ladish},
  \bibinfo{person}{Joshua Landau}, \bibinfo{person}{Kamal Ndousse},
  \bibinfo{person}{Kamile Lukosuite}, \bibinfo{person}{Liane Lovitt},
  \bibinfo{person}{Michael Sellitto}, \bibinfo{person}{Nelson Elhage},
  \bibinfo{person}{Nicholas Schiefer}, \bibinfo{person}{Noemi Mercado},
  \bibinfo{person}{Nova DasSarma}, \bibinfo{person}{Robert Lasenby},
  \bibinfo{person}{Robin Larson}, \bibinfo{person}{Sam Ringer},
  \bibinfo{person}{Scott Johnston}, \bibinfo{person}{Shauna Kravec},
  \bibinfo{person}{Sheer El~Showk}, \bibinfo{person}{Stanislav Fort},
  \bibinfo{person}{Tamera Lanham}, \bibinfo{person}{Timothy Telleen-Lawton},
  \bibinfo{person}{Tom Conerly}, \bibinfo{person}{Tom Henighan},
  \bibinfo{person}{Tristan Hume}, \bibinfo{person}{Samuel~R. Bowman},
  \bibinfo{person}{Zac Hatfield-Dodds}, \bibinfo{person}{Ben Mann},
  \bibinfo{person}{Dario Amodei}, \bibinfo{person}{Nicholas Joseph},
  \bibinfo{person}{Sam McCandlish}, \bibinfo{person}{Tom Brown}, {and}
  \bibinfo{person}{Jared Kaplan}.} \bibinfo{year}{2022}\natexlab{}.
\newblock \bibinfo{title}{Constitutional {AI}: Harmlessness from {AI}
  Feedback}.
\newblock \bibinfo{howpublished}{arXiv:2212.08073}.
\newblock


\bibitem[\protect\citeauthoryear{Bakker, Chadwick, Sheahan, Tessler,
  Campbell-Gillingham, Balaguer, McAleese, Glaese, Aslanides, Botvinick, and
  Summerfield}{Bakker et~al\mbox{.}}{2022}]%
        {bakker2022fine}
\bibfield{author}{\bibinfo{person}{Michiel~A. Bakker},
  \bibinfo{person}{Martin~J. Chadwick}, \bibinfo{person}{Hannah~R. Sheahan},
  \bibinfo{person}{Michael~Hnery Tessler}, \bibinfo{person}{Lucy
  Campbell-Gillingham}, \bibinfo{person}{Jan Balaguer}, \bibinfo{person}{Nat
  McAleese}, \bibinfo{person}{Amelia Glaese}, \bibinfo{person}{John Aslanides},
  \bibinfo{person}{Matthew~M. Botvinick}, {and} \bibinfo{person}{Christopher
  Summerfield}.} \bibinfo{year}{2022}\natexlab{}.
\newblock \showarticletitle{Fine-Tuning Language Models to Find Agreement Among
  Humans with Diverse Preferences}. In \bibinfo{booktitle}{\emph{Advances in
  Neural Information Processing Systems}}. \bibinfo{pages}{38176--38189}.
\newblock


\bibitem[\protect\citeauthoryear{Balaswamy}{Balaswamy}{2013}]%
        {balaswamy2013histories}
\bibfield{author}{\bibinfo{person}{Periaswamy Balaswamy}.}
  \bibinfo{year}{2013}\natexlab{}.
\newblock \bibinfo{title}{Histories From Below: The Condemned {A}halya, the
  Mortified {A}mba and the Oppressed {E}kalavya}.
\newblock \bibinfo{howpublished}{SSRN:3175708}.
\newblock


\bibitem[\protect\citeauthoryear{Bender, Gebru, McMillan-Major, and
  Shmitchell}{Bender et~al\mbox{.}}{2021}]%
        {BenderGMS2021}
\bibfield{author}{\bibinfo{person}{Emily~M. Bender}, \bibinfo{person}{Timnit
  Gebru}, \bibinfo{person}{Angelina McMillan-Major}, {and}
  \bibinfo{person}{Shmargaret Shmitchell}.} \bibinfo{year}{2021}\natexlab{}.
\newblock \showarticletitle{On the Dangers of Stochastic Parrots: Can Language
  Models Be Too Big?}. In \bibinfo{booktitle}{\emph{Proceedings of the ACM
  Conference on Fairness, Accountability, and Transparency}}.
  \bibinfo{pages}{610--623}.
\newblock


\bibitem[\protect\citeauthoryear{Benjamin}{Benjamin}{2022}]%
        {benjamin2022viral}
\bibfield{author}{\bibinfo{person}{Ruha Benjamin}.}
  \bibinfo{year}{2022}\natexlab{}.
\newblock \bibinfo{booktitle}{\emph{Viral Justice: How We Grow the World We
  Want}}.
\newblock \bibinfo{publisher}{Princeton University Press},
  \bibinfo{address}{Princeton, NJ, USA}.
\newblock


\bibitem[\protect\citeauthoryear{Benkler, Mosaphir, Friedman, Smart, and
  Schmer-Galunder}{Benkler et~al\mbox{.}}{2023}]%
        {BenklerMFSS2023}
\bibfield{author}{\bibinfo{person}{Noam Benkler}, \bibinfo{person}{Drisana
  Mosaphir}, \bibinfo{person}{Scott Friedman}, \bibinfo{person}{Andrew Smart},
  {and} \bibinfo{person}{Sonja Schmer-Galunder}.}
  \bibinfo{year}{2023}\natexlab{}.
\newblock \bibinfo{title}{Assessing {LLM}s for Moral Value Pluralism}.
\newblock \bibinfo{howpublished}{arXiv:2312.10075}.
\newblock


\bibitem[\protect\citeauthoryear{Benthall and Haynes}{Benthall and
  Haynes}{2019}]%
        {benthall2019racial}
\bibfield{author}{\bibinfo{person}{Sebastian Benthall} {and}
  \bibinfo{person}{Bruce~D. Haynes}.} \bibinfo{year}{2019}\natexlab{}.
\newblock \showarticletitle{Racial Categories in Machine Learning}. In
  \bibinfo{booktitle}{\emph{Proceedings of the Conference on Fairness,
  Accountability, and Transparency}}. \bibinfo{pages}{289--298}.
\newblock


\bibitem[\protect\citeauthoryear{Birhane}{Birhane}{2020}]%
        {birhane2020algorithmic}
\bibfield{author}{\bibinfo{person}{Abeba Birhane}.}
  \bibinfo{year}{2020}\natexlab{}.
\newblock \showarticletitle{Algorithmic Colonization of {A}frica}.
\newblock \bibinfo{journal}{\emph{SCRIPTed}} \bibinfo{volume}{17},
  \bibinfo{number}{2} (\bibinfo{date}{Aug.} \bibinfo{year}{2020}),
  \bibinfo{pages}{389--409}.
\newblock


\bibitem[\protect\citeauthoryear{Birhane}{Birhane}{2021}]%
        {birhane2021algorithmic}
\bibfield{author}{\bibinfo{person}{Abeba Birhane}.}
  \bibinfo{year}{2021}\natexlab{}.
\newblock \showarticletitle{Algorithmic Injustice: A Relational Ethics
  Approach}.
\newblock \bibinfo{journal}{\emph{Patterns}} \bibinfo{volume}{2},
  \bibinfo{number}{2} (\bibinfo{date}{Feb.} \bibinfo{year}{2021}),
  \bibinfo{pages}{100205}.
\newblock


\bibitem[\protect\citeauthoryear{Birhane, Ruane, Laurent, S.~Brown, Flowers,
  Ventresque, and L.~Dancy}{Birhane et~al\mbox{.}}{2022}]%
        {birhane2022forgotten}
\bibfield{author}{\bibinfo{person}{Abeba Birhane}, \bibinfo{person}{Elayne
  Ruane}, \bibinfo{person}{Thomas Laurent}, \bibinfo{person}{Matthew S.~Brown},
  \bibinfo{person}{Johnathan Flowers}, \bibinfo{person}{Anthony Ventresque},
  {and} \bibinfo{person}{Christopher L.~Dancy}.}
  \bibinfo{year}{2022}\natexlab{}.
\newblock \showarticletitle{The Forgotten Margins of {AI} Ethics}. In
  \bibinfo{booktitle}{\emph{Proceedings of the ACM Conference on Fairness,
  Accountability, and Transparency}}. \bibinfo{pages}{948--958}.
\newblock


\bibitem[\protect\citeauthoryear{Bommasani, Kapoor, Klyman, Longpre, Ramaswami,
  Zhang, Schaake, Ho, Narayanan, and Liang}{Bommasani et~al\mbox{.}}{2023}]%
        {Bommasani2023}
\bibfield{author}{\bibinfo{person}{Rishi Bommasani}, \bibinfo{person}{Sayash
  Kapoor}, \bibinfo{person}{Kevin Klyman}, \bibinfo{person}{Shayne Longpre},
  \bibinfo{person}{Ashwin Ramaswami}, \bibinfo{person}{Daniel Zhang},
  \bibinfo{person}{Marietje Schaake}, \bibinfo{person}{Daniel~E. Ho},
  \bibinfo{person}{Arvind Narayanan}, {and} \bibinfo{person}{Percy Liang}.}
  \bibinfo{year}{2023}\natexlab{}.
\newblock \bibinfo{booktitle}{\emph{Considerations for Governing Open
  Foundation Models}}.
\newblock \bibinfo{type}{Issue Brief}. \bibinfo{institution}{HAI Policy \&
  Society}.
\newblock


\bibitem[\protect\citeauthoryear{Bornstein, Appenzeller, and Casado}{Bornstein
  et~al\mbox{.}}{2023}]%
        {BornsteinAC2023}
\bibfield{author}{\bibinfo{person}{Matt Bornstein}, \bibinfo{person}{Guido
  Appenzeller}, {and} \bibinfo{person}{Martin Casado}.}
  \bibinfo{year}{2023}\natexlab{}.
\newblock \bibinfo{title}{Who Owns the Generative AI Platform?}
\newblock
  \bibinfo{howpublished}{https://a16z.com/who-owns-the-generative-ai-platform/}.
\newblock


\bibitem[\protect\citeauthoryear{Bouneffouf}{Bouneffouf}{2023}]%
        {Bouneffouf2023}
\bibfield{author}{\bibinfo{person}{Djallel Bouneffouf}.}
  \bibinfo{year}{2023}\natexlab{}.
\newblock \bibinfo{booktitle}{\emph{Multi-Armed Bandit Problem and
  Application}}.
\newblock \bibinfo{publisher}{Independently Published}.
\newblock


\bibitem[\protect\citeauthoryear{Buolamwini and Gebru}{Buolamwini and
  Gebru}{2018}]%
        {buolamwini2018gender}
\bibfield{author}{\bibinfo{person}{Joy Buolamwini} {and}
  \bibinfo{person}{Timnit Gebru}.} \bibinfo{year}{2018}\natexlab{}.
\newblock \showarticletitle{Gender Shades: Intersectional Accuracy Disparities
  in Commercial Gender Classification}. In
  \bibinfo{booktitle}{\emph{Proceedings of the Conference on Fairness,
  Accountability and Transparency}}. \bibinfo{pages}{77--91}.
\newblock


\bibitem[\protect\citeauthoryear{Carpenter}{Carpenter}{2005}]%
        {liu2006conceptions}
\bibfield{author}{\bibinfo{person}{Amber Carpenter}.}
  \bibinfo{year}{2005}\natexlab{}.
\newblock \showarticletitle{Questioning {K\d{r}\d{s}\d{n}}a’s {K}antianism}.
\newblock In \bibinfo{booktitle}{\emph{Conceptions of Virtue: East and West}},
  \bibfield{editor}{\bibinfo{person}{Kim-Chong Chong} {and}
  \bibinfo{person}{Yuli Liu}} (Eds.). \bibinfo{publisher}{Marshall Cavendish},
  \bibinfo{pages}{80--99}.
\newblock


\bibitem[\protect\citeauthoryear{Casper, Davies, Shi, Gilbert, Scheurer, Rando,
  Freedman, Korbak, Lindner, Freire, Wang, Marks, Segerie, Carroll, Peng,
  Christoffersen, Damani, Slocum, Anwar, Siththaranjan, Nadeau, Michaud, Pfau,
  Krasheninnikov, Chen, Langosco, Hase, Bıyık, Dragan, Krueger, Sadigh, and
  Hadfield-Menell}{Casper et~al\mbox{.}}{2023}]%
        {casper2023open}
\bibfield{author}{\bibinfo{person}{Stephen Casper}, \bibinfo{person}{Xander
  Davies}, \bibinfo{person}{Claudia Shi}, \bibinfo{person}{Thomas~Krendl
  Gilbert}, \bibinfo{person}{J{\'e}r{\'e}my Scheurer}, \bibinfo{person}{Javier
  Rando}, \bibinfo{person}{Rachel Freedman}, \bibinfo{person}{Tomasz Korbak},
  \bibinfo{person}{David Lindner}, \bibinfo{person}{Pedro Freire},
  \bibinfo{person}{Tony Wang}, \bibinfo{person}{Samuel Marks},
  \bibinfo{person}{Charbel-Raphaël Segerie}, \bibinfo{person}{Micah Carroll},
  \bibinfo{person}{Andi Peng}, \bibinfo{person}{Phillip Christoffersen},
  \bibinfo{person}{Mehul Damani}, \bibinfo{person}{Stewart Slocum},
  \bibinfo{person}{Usman Anwar}, \bibinfo{person}{Anand Siththaranjan},
  \bibinfo{person}{Max Nadeau}, \bibinfo{person}{Eric~J. Michaud},
  \bibinfo{person}{Jacob Pfau}, \bibinfo{person}{Dmitrii Krasheninnikov},
  \bibinfo{person}{Xin Chen}, \bibinfo{person}{Lauro Langosco},
  \bibinfo{person}{Peter Hase}, \bibinfo{person}{Erdem Bıyık},
  \bibinfo{person}{Anca Dragan}, \bibinfo{person}{David Krueger},
  \bibinfo{person}{Dorsa Sadigh}, {and} \bibinfo{person}{Dylan
  Hadfield-Menell}.} \bibinfo{year}{2023}\natexlab{}.
\newblock \bibinfo{title}{Open Problems and Fundamental Limitations of
  Reinforcement Learning from Human Feedback}.
\newblock \bibinfo{howpublished}{arXiv:2307.15217}.
\newblock


\bibitem[\protect\citeauthoryear{Cave}{Cave}{2020}]%
        {cave2020problem}
\bibfield{author}{\bibinfo{person}{Stephen Cave}.}
  \bibinfo{year}{2020}\natexlab{}.
\newblock \showarticletitle{The Problem with Intelligence: Its Value-Laden
  History and the Future of {AI}}. In \bibinfo{booktitle}{\emph{Proceedings of
  the AAAI/ACM Conference on AI, Ethics, and Society}}.
  \bibinfo{pages}{29--35}.
\newblock


\bibitem[\protect\citeauthoryear{Chan, Hall, Piron, Tandon, and Williams}{Chan
  et~al\mbox{.}}{2020}]%
        {chan2020open}
\bibfield{author}{\bibinfo{person}{Leslie Chan}, \bibinfo{person}{Budd Hall},
  \bibinfo{person}{Florence Piron}, \bibinfo{person}{Rajesh Tandon}, {and}
  \bibinfo{person}{Lorna Williams}.} \bibinfo{year}{2020}\natexlab{}.
\newblock \bibinfo{booktitle}{\emph{Open Science Beyond Open Access: For and
  With Communities: A Step Towards the Decolonization of Knowledge}}.
\newblock \bibinfo{type}{{T}echnical {R}eport}. \bibinfo{institution}{Canadian
  Commission for UNESCO}.
\newblock


\bibitem[\protect\citeauthoryear{Costanza-Chock}{Costanza-Chock}{2020}]%
        {costanza2020design}
\bibfield{author}{\bibinfo{person}{Sasha Costanza-Chock}.}
  \bibinfo{year}{2020}\natexlab{}.
\newblock \bibinfo{booktitle}{\emph{Design Justice: Community-Led Practices to
  Build the Worlds We Need}}.
\newblock \bibinfo{publisher}{MIT Press}, \bibinfo{address}{Cambridge, MA,
  USA}.
\newblock


\bibitem[\protect\citeauthoryear{Crawford}{Crawford}{2021}]%
        {crawford2021atlas}
\bibfield{author}{\bibinfo{person}{Kate Crawford}.}
  \bibinfo{year}{2021}\natexlab{}.
\newblock \bibinfo{booktitle}{\emph{The Atlas of {AI}: Power, Politics, and the
  Planetary Costs of Artificial Intelligence}}.
\newblock \bibinfo{publisher}{Yale University Press}, \bibinfo{address}{New
  Haven, CT, USA}.
\newblock


\bibitem[\protect\citeauthoryear{Das, {\O}sterlund, and Semaan}{Das
  et~al\mbox{.}}{2021}]%
        {das2021jol}
\bibfield{author}{\bibinfo{person}{Dipto Das}, \bibinfo{person}{Carsten
  {\O}sterlund}, {and} \bibinfo{person}{Bryan Semaan}.}
  \bibinfo{year}{2021}\natexlab{}.
\newblock \showarticletitle{"Jol" or "Pani"?: How Does Governance Shape a
  Platform's Identity?}
\newblock \bibinfo{journal}{\emph{Proceedings of the ACM on Human-Computer
  Interaction}} \bibinfo{volume}{5}, \bibinfo{number}{CSCW2}
  (\bibinfo{year}{2021}), \bibinfo{pages}{473}.
\newblock


\bibitem[\protect\citeauthoryear{Das}{Das}{2014}]%
        {das2014valmiki}
\bibfield{author}{\bibinfo{person}{Rumpa Das}.}
  \bibinfo{year}{2014}\natexlab{}.
\newblock \showarticletitle{{V}almiki {P}ratibha (The Genius of {V}almiki): A
  Study in Genius}.
\newblock In \bibinfo{booktitle}{\emph{The Politics and Reception of
  Rabindranath Tagore's Drama}}. \bibinfo{publisher}{Routledge},
  \bibinfo{address}{New York, NY, USA}, \bibinfo{pages}{103--112}.
\newblock


\bibitem[\protect\citeauthoryear{de~Sousa~Santos}{de~Sousa~Santos}{2011}]%
        {de2011epistemologias}
\bibfield{author}{\bibinfo{person}{Boaventura de Sousa~Santos}.}
  \bibinfo{year}{2011}\natexlab{}.
\newblock \showarticletitle{Epistemolog{\'\i}as del Sur}.
\newblock \bibinfo{journal}{\emph{Utop{\'\i}a y Praxis Latinoamericana}}
  \bibinfo{volume}{16}, \bibinfo{number}{54} (\bibinfo{date}{July--Sept.}
  \bibinfo{year}{2011}), \bibinfo{pages}{17--39}.
\newblock


\bibitem[\protect\citeauthoryear{Dev, Goyal, Tewari, Dave, and Prabhakaran}{Dev
  et~al\mbox{.}}{2023}]%
        {dev2023building}
\bibfield{author}{\bibinfo{person}{Sunipa Dev}, \bibinfo{person}{Jaya Goyal},
  \bibinfo{person}{Dinesh Tewari}, \bibinfo{person}{Shachi Dave}, {and}
  \bibinfo{person}{Vinodkumar Prabhakaran}.} \bibinfo{year}{2023}\natexlab{}.
\newblock \bibinfo{title}{Building Socio-culturally Inclusive Stereotype
  Resources with Community Engagement}.
\newblock \bibinfo{howpublished}{arXiv:2307.10514}.
\newblock


\bibitem[\protect\citeauthoryear{Dhand}{Dhand}{2002}]%
        {dhand2002dharma}
\bibfield{author}{\bibinfo{person}{Arti Dhand}.}
  \bibinfo{year}{2002}\natexlab{}.
\newblock \showarticletitle{The Dharma of Ethics, The Ethics of Dharma:
  Quizzing the Ideals of {H}induism}.
\newblock \bibinfo{journal}{\emph{Journal of Religious Ethics}}
  \bibinfo{volume}{30}, \bibinfo{number}{3} (\bibinfo{date}{Fall}
  \bibinfo{year}{2002}), \bibinfo{pages}{347--372}.
\newblock


\bibitem[\protect\citeauthoryear{Diagne}{Diagne}{2011}]%
        {diagne2023african}
\bibfield{author}{\bibinfo{person}{Souleymane~Bachir Diagne}.}
  \bibinfo{year}{2011}\natexlab{}.
\newblock \bibinfo{booktitle}{\emph{African Art as Philosophy: Senghor,
  Bergson, and the Idea of Negritude}}.
\newblock \bibinfo{publisher}{Seagull Books}, \bibinfo{address}{London, UK}.
\newblock


\bibitem[\protect\citeauthoryear{Dirks}{Dirks}{2002}]%
        {dirks2002castes}
\bibfield{author}{\bibinfo{person}{Nicholas~B. Dirks}.}
  \bibinfo{year}{2002}\natexlab{}.
\newblock \bibinfo{booktitle}{\emph{Castes of Mind: Colonialism and the Making
  of Modern {I}ndia}}.
\newblock \bibinfo{publisher}{Princeton University Press},
  \bibinfo{address}{Princeton, NJ, USA}.
\newblock


\bibitem[\protect\citeauthoryear{Divakaran, Sridhar, and Srinivasan}{Divakaran
  et~al\mbox{.}}{2023}]%
        {divakaran2023broadening}
\bibfield{author}{\bibinfo{person}{Ajay Divakaran}, \bibinfo{person}{Aparna
  Sridhar}, {and} \bibinfo{person}{Ramya Srinivasan}.}
  \bibinfo{year}{2023}\natexlab{}.
\newblock \showarticletitle{Broadening {AI} Ethics Narratives: An {I}ndic Art
  View}. In \bibinfo{booktitle}{\emph{Proceedings of the ACM Conference on
  Fairness, Accountability, and Transparency}}. \bibinfo{pages}{2--11}.
\newblock


\bibitem[\protect\citeauthoryear{Dognin, Rios, Luss, Padhi, Riemer, Liu,
  Sattigeri, Nagireddy, Varshney, and Bouneffouf}{Dognin et~al\mbox{.}}{2024}]%
        {dognin2024contextual}
\bibfield{author}{\bibinfo{person}{Pierre Dognin}, \bibinfo{person}{Jesus
  Rios}, \bibinfo{person}{Ronny Luss}, \bibinfo{person}{Inkit Padhi},
  \bibinfo{person}{Matthew~D. Riemer}, \bibinfo{person}{Miao Liu},
  \bibinfo{person}{Prasanna Sattigeri}, \bibinfo{person}{Manish Nagireddy},
  \bibinfo{person}{Kush~R. Varshney}, {and} \bibinfo{person}{Djallel
  Bouneffouf}.} \bibinfo{year}{2024}\natexlab{}.
\newblock \showarticletitle{Contextual Moral Value Alignment Through
  Context-Based Aggregation}.
\newblock In \bibinfo{booktitle}{\emph{Proceedings of the International Joint
  Conference on Artificial Intelligence}}.
\newblock


\bibitem[\protect\citeauthoryear{Dunford}{Dunford}{2017}]%
        {dunford2017toward}
\bibfield{author}{\bibinfo{person}{Robin Dunford}.}
  \bibinfo{year}{2017}\natexlab{}.
\newblock \showarticletitle{Toward a Decolonial Global Ethics}.
\newblock \bibinfo{journal}{\emph{Journal of Global Ethics}}
  \bibinfo{volume}{13}, \bibinfo{number}{3} (\bibinfo{date}{March}
  \bibinfo{year}{2017}), \bibinfo{pages}{380--397}.
\newblock


\bibitem[\protect\citeauthoryear{Durmus, Nyugen, Liao, Schiefer, Askell,
  Bakhtin, Chen, Hatfield-Dodds, Hernandez, Joseph, Lovitt, McCandlish, Sikder,
  Tamkin, Thamkul, Kaplan, Clark, and Ganguli}{Durmus et~al\mbox{.}}{2023}]%
        {Durmus2023}
\bibfield{author}{\bibinfo{person}{Esin Durmus}, \bibinfo{person}{Karina
  Nyugen}, \bibinfo{person}{Thomas~I. Liao}, \bibinfo{person}{Nicholas
  Schiefer}, \bibinfo{person}{Amanda Askell}, \bibinfo{person}{Anton Bakhtin},
  \bibinfo{person}{Carol Chen}, \bibinfo{person}{Zac Hatfield-Dodds},
  \bibinfo{person}{Danny Hernandez}, \bibinfo{person}{Nicholas Joseph},
  \bibinfo{person}{Liane Lovitt}, \bibinfo{person}{Sam McCandlish},
  \bibinfo{person}{Orowa Sikder}, \bibinfo{person}{Alex Tamkin},
  \bibinfo{person}{Janel Thamkul}, \bibinfo{person}{Jared Kaplan},
  \bibinfo{person}{Jack Clark}, {and} \bibinfo{person}{Deep Ganguli}.}
  \bibinfo{year}{2023}\natexlab{}.
\newblock \bibinfo{title}{Towards Measuring the Representation of Subjective
  Global Opinions in Language Models}.
\newblock \bibinfo{howpublished}{arXiv:2306.16388}.
\newblock


\bibitem[\protect\citeauthoryear{Dutta, Ramasubramanian, Barrett, Elers,
  Sarwatay, Raghunath, Kaur, Dutta, Jayan, Rahman, Tallam, Roy, Falnikar,
  Johnson, Mandal, Dutta, Basnyat, Soriano, Pavarala, Sreekumar, Ganesh, Pandi,
  and Zapata}{Dutta et~al\mbox{.}}{2021}]%
        {dutta2021decolonizing}
\bibfield{author}{\bibinfo{person}{Mohan Dutta}, \bibinfo{person}{Srividya
  Ramasubramanian}, \bibinfo{person}{Mereana Barrett},
  \bibinfo{person}{Christine Elers}, \bibinfo{person}{Devina Sarwatay},
  \bibinfo{person}{Preeti Raghunath}, \bibinfo{person}{Satveer Kaur},
  \bibinfo{person}{Debalina Dutta}, \bibinfo{person}{Pooja Jayan},
  \bibinfo{person}{Mahbubur Rahman}, \bibinfo{person}{Edwin Tallam},
  \bibinfo{person}{Sudeshna Roy}, \bibinfo{person}{Ashwini Falnikar},
  \bibinfo{person}{Gayle~Moana Johnson}, \bibinfo{person}{Indranil Mandal},
  \bibinfo{person}{Uttaran Dutta}, \bibinfo{person}{Iccha Basnyat},
  \bibinfo{person}{Cheryll Soriano}, \bibinfo{person}{Vinod Pavarala},
  \bibinfo{person}{T.~T. Sreekumar}, \bibinfo{person}{Shiv Ganesh},
  \bibinfo{person}{Asha~Rathina Pandi}, {and} \bibinfo{person}{Dazzelyn
  Zapata}.} \bibinfo{year}{2021}\natexlab{}.
\newblock \showarticletitle{Decolonizing Open Science: {S}outhern
  Interventions}.
\newblock \bibinfo{journal}{\emph{Journal of Communication}}
  \bibinfo{volume}{71}, \bibinfo{number}{5} (\bibinfo{date}{Sept.}
  \bibinfo{year}{2021}), \bibinfo{pages}{803--826}.
\newblock


\bibitem[\protect\citeauthoryear{Dzieza}{Dzieza}{2023}]%
        {Dzieza2023}
\bibfield{author}{\bibinfo{person}{Josh Dzieza}.}
  \bibinfo{year}{2023}\natexlab{}.
\newblock \showarticletitle{{AI} Is a Lot of Work}.
\newblock \bibinfo{journal}{\emph{New York}} (\bibinfo{date}{June}
  \bibinfo{year}{2023}).
\newblock


\bibitem[\protect\citeauthoryear{D’Cruz, Kidder, and Varshney}{D’Cruz
  et~al\mbox{.}}{2022}]%
        {d2022empathy}
\bibfield{author}{\bibinfo{person}{Jason~R. D’Cruz}, \bibinfo{person}{William
  Kidder}, {and} \bibinfo{person}{Kush~R. Varshney}.}
  \bibinfo{year}{2022}\natexlab{}.
\newblock \showarticletitle{The Empathy Gap: Why {AI} Can Forecast Behavior But
  Cannot Assess Trustworthiness}. In \bibinfo{booktitle}{\emph{Proceedings of
  the AAAI Fall Symposium Series Symposium on Thinking Fast and Slow and Other
  Cognitive Theories in AI}}. \bibinfo{pages}{9}.
\newblock


\bibitem[\protect\citeauthoryear{Ehsan, Singh, Metcalf, and Riedl}{Ehsan
  et~al\mbox{.}}{2022}]%
        {ehsan2022algorithmic}
\bibfield{author}{\bibinfo{person}{Upol Ehsan}, \bibinfo{person}{Ranjit Singh},
  \bibinfo{person}{Jacob Metcalf}, {and} \bibinfo{person}{Mark Riedl}.}
  \bibinfo{year}{2022}\natexlab{}.
\newblock \showarticletitle{The Algorithmic Imprint}. In
  \bibinfo{booktitle}{\emph{Proceedings of the ACM Conference on Fairness,
  Accountability, and Transparency}}. \bibinfo{pages}{1305--1317}.
\newblock


\bibitem[\protect\citeauthoryear{Erman and M{\"{o}}ller}{Erman and
  M{\"{o}}ller}{2018}]%
        {ErmanM2018}
\bibfield{author}{\bibinfo{person}{Eva Erman} {and} \bibinfo{person}{Niklas
  M{\"{o}}ller}.} \bibinfo{year}{2018}\natexlab{}.
\newblock \showarticletitle{The Interdependence of Risk and Moral Theory}.
\newblock \bibinfo{journal}{\emph{Ethical Theory and Moral Practice}}
  \bibinfo{volume}{21}, \bibinfo{number}{2} (\bibinfo{date}{March}
  \bibinfo{year}{2018}), \bibinfo{pages}{207--216}.
\newblock


\bibitem[\protect\citeauthoryear{{\SortNoop{Es}}van~Es, Everts, and
  Muis}{{\SortNoop{Es}}van~Es et~al\mbox{.}}{2021}]%
        {van_Es_Everts_Muis_2021}
\bibfield{author}{\bibinfo{person}{Karin {\SortNoop{Es}}van~Es},
  \bibinfo{person}{Daniel Everts}, {and} \bibinfo{person}{Iris Muis}.}
  \bibinfo{year}{2021}\natexlab{}.
\newblock \showarticletitle{Gendered Language and Employment Web Sites: How
  Search Algorithms Can Cause Allocative Harm}.
\newblock \bibinfo{journal}{\emph{First Monday}} \bibinfo{volume}{26},
  \bibinfo{number}{8} (\bibinfo{date}{July} \bibinfo{year}{2021}).
\newblock


\bibitem[\protect\citeauthoryear{Feng, Park, Liu, and Tsvetkov}{Feng
  et~al\mbox{.}}{2023}]%
        {Feng2023}
\bibfield{author}{\bibinfo{person}{Shangbin Feng}, \bibinfo{person}{Chan~Young
  Park}, \bibinfo{person}{Yuhan Liu}, {and} \bibinfo{person}{Yulia Tsvetkov}.}
  \bibinfo{year}{2023}\natexlab{}.
\newblock \bibinfo{title}{From Pretraining Data to Language Models to
  Downstream Tasks: Tracking the Trails of Political Biases Leading to Unfair
  {NLP} Models}.
\newblock \bibinfo{howpublished}{arXiv:2305.08283}.
\newblock


\bibitem[\protect\citeauthoryear{Flanagan}{Flanagan}{2016}]%
        {flanagan2016geography}
\bibfield{author}{\bibinfo{person}{Owen Flanagan}.}
  \bibinfo{year}{2016}\natexlab{}.
\newblock \bibinfo{booktitle}{\emph{The Geography of Morals: Varieties of Moral
  Possibility}}.
\newblock \bibinfo{publisher}{Oxford University Press}, \bibinfo{address}{New
  York, NY, USA}.
\newblock


\bibitem[\protect\citeauthoryear{Frawley}{Frawley}{2023}]%
        {Frawley2023}
\bibfield{author}{\bibinfo{person}{David Frawley}.}
  \bibinfo{year}{2023}\natexlab{}.
\newblock
  \bibinfo{howpublished}{https://twitter.com/davidfrawleyved/status/1681689995554476033?s=20}.
\newblock


\bibitem[\protect\citeauthoryear{Gabriel}{Gabriel}{2020}]%
        {Gabriel2020}
\bibfield{author}{\bibinfo{person}{Iason Gabriel}.}
  \bibinfo{year}{2020}\natexlab{}.
\newblock \showarticletitle{Artificial Intelligence, Values, and Alignment}.
\newblock \bibinfo{journal}{\emph{Minds and Machines}} \bibinfo{volume}{30},
  \bibinfo{number}{3} (\bibinfo{date}{Sept.} \bibinfo{year}{2020}),
  \bibinfo{pages}{411--437}.
\newblock


\bibitem[\protect\citeauthoryear{Gottheil}{Gottheil}{1977}]%
        {e53927b8-0a79-3d19-93b7-95642d44b953}
\bibfield{author}{\bibinfo{person}{Fred~M. Gottheil}.}
  \bibinfo{year}{1977}\natexlab{}.
\newblock \showarticletitle{On an Economic Theory of Colonialism}.
\newblock \bibinfo{journal}{\emph{Journal of Economic Issues}}
  \bibinfo{volume}{11}, \bibinfo{number}{1} (\bibinfo{date}{March}
  \bibinfo{year}{1977}), \bibinfo{pages}{83--102}.
\newblock


\bibitem[\protect\citeauthoryear{Gray and Witt}{Gray and Witt}{2021}]%
        {Gray_Witt_2021}
\bibfield{author}{\bibinfo{person}{Joanne Gray} {and} \bibinfo{person}{Alice
  Witt}.} \bibinfo{year}{2021}\natexlab{}.
\newblock \showarticletitle{A Feminist Data Ethics of Care for Machine
  Learning: The What, Why, Who and How}.
\newblock \bibinfo{journal}{\emph{First Monday}} \bibinfo{volume}{26},
  \bibinfo{number}{12} (\bibinfo{date}{Dec.} \bibinfo{year}{2021}).
\newblock


\bibitem[\protect\citeauthoryear{Gray and Suri}{Gray and Suri}{2019}]%
        {gray2019ghost}
\bibfield{author}{\bibinfo{person}{Mary~L. Gray} {and}
  \bibinfo{person}{Siddharth Suri}.} \bibinfo{year}{2019}\natexlab{}.
\newblock \bibinfo{booktitle}{\emph{Ghost Work: How to Stop {S}ilicon {V}alley
  from Building a New Global Underclass}}.
\newblock \bibinfo{publisher}{Houghton Mifflin Harcourt}, \bibinfo{address}{New
  York, NY, USA}.
\newblock


\bibitem[\protect\citeauthoryear{Green}{Green}{2019}]%
        {green2019good}
\bibfield{author}{\bibinfo{person}{Ben Green}.}
  \bibinfo{year}{2019}\natexlab{}.
\newblock \showarticletitle{``Good'' Isn't Good Enough}. In
  \bibinfo{booktitle}{\emph{Proceedings of the NeurIPS AI for Social Good
  Workshop}}.
\newblock


\bibitem[\protect\citeauthoryear{Greene, Dhurandhar, and Shmueli}{Greene
  et~al\mbox{.}}{2023}]%
        {greene2023atomist}
\bibfield{author}{\bibinfo{person}{Travis Greene}, \bibinfo{person}{Amit
  Dhurandhar}, {and} \bibinfo{person}{Galit Shmueli}.}
  \bibinfo{year}{2023}\natexlab{}.
\newblock \showarticletitle{Atomist or Holist? {A} Diagnosis and Vision for
  More Productive Interdisciplinary {AI} Ethics Dialogue}.
\newblock \bibinfo{journal}{\emph{Patterns}} \bibinfo{volume}{4},
  \bibinfo{number}{1} (\bibinfo{date}{Jan.} \bibinfo{year}{2023}),
  \bibinfo{pages}{100652}.
\newblock


\bibitem[\protect\citeauthoryear{Grosfoguel}{Grosfoguel}{2013}]%
        {grosfoguel2013structure}
\bibfield{author}{\bibinfo{person}{Ram{\'o}n Grosfoguel}.}
  \bibinfo{year}{2013}\natexlab{}.
\newblock \showarticletitle{The Structure of Knowledge in Westernised
  Universities: Epistemic Racism/Sexism and the Four Genocides/Epistemicides}.
\newblock \bibinfo{journal}{\emph{Human Architecture: Journal of the Sociology
  of Self-Knowledge}} \bibinfo{volume}{11}, \bibinfo{number}{1}
  (\bibinfo{date}{Fall} \bibinfo{year}{2013}), \bibinfo{pages}{73--90}.
\newblock


\bibitem[\protect\citeauthoryear{Haemmerl, Deiseroth, Schramowski,
  Libovick{\'y}, Rothkopf, Fraser, and Kersting}{Haemmerl
  et~al\mbox{.}}{2023}]%
        {haemmerl-etal-2023-speaking}
\bibfield{author}{\bibinfo{person}{Katharina Haemmerl}, \bibinfo{person}{Bjoern
  Deiseroth}, \bibinfo{person}{Patrick Schramowski},
  \bibinfo{person}{Jind{\v{r}}ich Libovick{\'y}}, \bibinfo{person}{Constantin
  Rothkopf}, \bibinfo{person}{Alexander Fraser}, {and}
  \bibinfo{person}{Kristian Kersting}.} \bibinfo{year}{2023}\natexlab{}.
\newblock \showarticletitle{Speaking Multiple Languages Affects the Moral Bias
  of Language Models}. In \bibinfo{booktitle}{\emph{Findings of the Association
  for Computational Linguistics}}. \bibinfo{pages}{2137--2156}.
\newblock


\bibitem[\protect\citeauthoryear{Hall and Tandon}{Hall and Tandon}{2017}]%
        {hall2017decolonization}
\bibfield{author}{\bibinfo{person}{Budd~L. Hall} {and} \bibinfo{person}{Rajesh
  Tandon}.} \bibinfo{year}{2017}\natexlab{}.
\newblock \showarticletitle{Decolonization of Knowledge, Epistemicide,
  Participatory Research and Higher Education}.
\newblock \bibinfo{journal}{\emph{Research for All}} \bibinfo{volume}{1},
  \bibinfo{number}{1} (\bibinfo{date}{Jan.} \bibinfo{year}{2017}),
  \bibinfo{pages}{6--19}.
\newblock


\bibitem[\protect\citeauthoryear{Hao}{Hao}{2022}]%
        {Hao2022}
\bibfield{author}{\bibinfo{person}{Karen Hao}.}
  \bibinfo{year}{2022}\natexlab{}.
\newblock \showarticletitle{Artificial Intelligence is Creating a New Colonial
  World Order}.
\newblock \bibinfo{journal}{\emph{MIT Technology Review}}
  (\bibinfo{date}{April} \bibinfo{year}{2022}).
\newblock


\bibitem[\protect\citeauthoryear{Hardt and Negri}{Hardt and Negri}{2001}]%
        {hardt2001empire}
\bibfield{author}{\bibinfo{person}{Michael Hardt} {and}
  \bibinfo{person}{Antonio Negri}.} \bibinfo{year}{2001}\natexlab{}.
\newblock \bibinfo{booktitle}{\emph{Empire}}.
\newblock \bibinfo{publisher}{Harvard University Press},
  \bibinfo{address}{Cambridge, MA, USA}.
\newblock


\bibitem[\protect\citeauthoryear{Harris}{Harris}{2024}]%
        {Harris2024}
\bibfield{author}{\bibinfo{person}{David~Evan Harris}.}
  \bibinfo{year}{2024}\natexlab{}.
\newblock \showarticletitle{Open-Source {AI} Is Uniquely Dangerous}.
\newblock \bibinfo{journal}{\emph{IEEE Spectrum}} (\bibinfo{year}{2024}).
\newblock


\bibitem[\protect\citeauthoryear{Hassan}{Hassan}{2023}]%
        {hassan2023governing}
\bibfield{author}{\bibinfo{person}{Yousif Hassan}.}
  \bibinfo{year}{2023}\natexlab{}.
\newblock \showarticletitle{Governing Algorithms from the {S}outh: A Case Study
  of {AI} Development in {A}frica}.
\newblock \bibinfo{journal}{\emph{AI \& Society}} \bibinfo{volume}{38},
  \bibinfo{number}{4} (\bibinfo{year}{2023}), \bibinfo{pages}{1429--1442}.
\newblock


\bibitem[\protect\citeauthoryear{Hlabangane}{Hlabangane}{2021}]%
        {hlabangane2021underside}
\bibfield{author}{\bibinfo{person}{Nokuthula Hlabangane}.}
  \bibinfo{year}{2021}\natexlab{}.
\newblock \showarticletitle{The Underside of Modern Knowledge: An Epistemic
  Break from Western Science}.
\newblock In \bibinfo{booktitle}{\emph{Decolonising the Human: Reflections from
  {A}frica on Difference and Oppression}},
  \bibfield{editor}{\bibinfo{person}{Melissa Steyn} {and}
  \bibinfo{person}{William Mpofu}} (Eds.). \bibinfo{publisher}{Wits University
  Press}, \bibinfo{address}{Johannesburg, South Africa},
  \bibinfo{pages}{164--185}.
\newblock


\bibitem[\protect\citeauthoryear{Hu, Shen, Wallis, Allen-Zhu, Li, Wang, Wang,
  and Chen}{Hu et~al\mbox{.}}{2022}]%
        {hu2021lora}
\bibfield{author}{\bibinfo{person}{Edward Hu}, \bibinfo{person}{Yelong Shen},
  \bibinfo{person}{Phillip Wallis}, \bibinfo{person}{Zeyuan Allen-Zhu},
  \bibinfo{person}{Yuanzhi Li}, \bibinfo{person}{Shean Wang},
  \bibinfo{person}{Lu Wang}, {and} \bibinfo{person}{Weizhu Chen}.}
  \bibinfo{year}{2022}\natexlab{}.
\newblock \showarticletitle{{LoRA}: Low-Rank Adaptation of Large Language
  Models}. In \bibinfo{booktitle}{\emph{Proceedings of the International
  Conference on Learning Representations}}.
\newblock


\bibitem[\protect\citeauthoryear{Huang, Liu, Lin, Pang, Du, and Lin}{Huang
  et~al\mbox{.}}{2023}]%
        {huang2023lorahub}
\bibfield{author}{\bibinfo{person}{Chengsong Huang}, \bibinfo{person}{Qian
  Liu}, \bibinfo{person}{Bill~Yuchen Lin}, \bibinfo{person}{Tianyu Pang},
  \bibinfo{person}{Chao Du}, {and} \bibinfo{person}{Min Lin}.}
  \bibinfo{year}{2023}\natexlab{}.
\newblock \bibinfo{title}{{LoraHub}: Efficient Cross-Task Generalization via
  Dynamic {LoRA} Composition}.
\newblock \bibinfo{howpublished}{arXiv:2307.13269}.
\newblock


\bibitem[\protect\citeauthoryear{Igboin}{Igboin}{2011}]%
        {igboin2011colonialism}
\bibfield{author}{\bibinfo{person}{Benson~O. Igboin}.}
  \bibinfo{year}{2011}\natexlab{}.
\newblock \showarticletitle{Colonialism and {A}frican Cultural Values}.
\newblock \bibinfo{journal}{\emph{African Journal of History and Culture}}
  \bibinfo{volume}{3}, \bibinfo{number}{6} (\bibinfo{date}{July}
  \bibinfo{year}{2011}), \bibinfo{pages}{96--103}.
\newblock


\bibitem[\protect\citeauthoryear{Jackson}{Jackson}{1972}]%
        {jackson1972aspects}
\bibfield{author}{\bibinfo{person}{Michael Jackson}.}
  \bibinfo{year}{1972}\natexlab{}.
\newblock \showarticletitle{Aspects of Symbolism and Composition in {M}aori
  Art}.
\newblock \bibinfo{journal}{\emph{Bijdragen tot de Taal-, Land- en
  Volkenkunde}} \bibinfo{volume}{128}, \bibinfo{number}{1}
  (\bibinfo{date}{Jan.} \bibinfo{year}{1972}), \bibinfo{pages}{33--80}.
\newblock


\bibitem[\protect\citeauthoryear{Jang, Kim, Lin, Wang, Hessel, Zettlemoyer,
  Hajishirzi, Choi, and Ammanabrolu}{Jang et~al\mbox{.}}{2023}]%
        {jang2023personalized}
\bibfield{author}{\bibinfo{person}{Joel Jang}, \bibinfo{person}{Seungone Kim},
  \bibinfo{person}{Bill~Yuchen Lin}, \bibinfo{person}{Yizhong Wang},
  \bibinfo{person}{Jack Hessel}, \bibinfo{person}{Luke Zettlemoyer},
  \bibinfo{person}{Hannaneh Hajishirzi}, \bibinfo{person}{Yejin Choi}, {and}
  \bibinfo{person}{Prithviraj Ammanabrolu}.} \bibinfo{year}{2023}\natexlab{}.
\newblock \bibinfo{title}{Personalized Soups: Personalized Large Language Model
  Alignment via Post-Hoc Parameter Merging}.
\newblock \bibinfo{howpublished}{arXiv:2310.11564}.
\newblock


\bibitem[\protect\citeauthoryear{Ji, Qiu, Chen, Zhang, Lou, Wang, Duan, He,
  Zhou, Zhang, Zeng, Ng, Dai, Pan, O'Gara, Lei, Xu, Tse, Fu, McAleer, Yang,
  Wang, Zhu, Guo, and Gao}{Ji et~al\mbox{.}}{2023}]%
        {ji2023ai}
\bibfield{author}{\bibinfo{person}{Jiaming Ji}, \bibinfo{person}{Tianyi Qiu},
  \bibinfo{person}{Boyuan Chen}, \bibinfo{person}{Borong Zhang},
  \bibinfo{person}{Hantao Lou}, \bibinfo{person}{Kaile Wang},
  \bibinfo{person}{Yawen Duan}, \bibinfo{person}{Zhonghao He},
  \bibinfo{person}{Jiayi Zhou}, \bibinfo{person}{Zhaowei Zhang},
  \bibinfo{person}{Fanzhi Zeng}, \bibinfo{person}{Kwan~Yee Ng},
  \bibinfo{person}{Juntao Dai}, \bibinfo{person}{Xuehai Pan},
  \bibinfo{person}{Aidan O'Gara}, \bibinfo{person}{Yingshan Lei},
  \bibinfo{person}{Hua Xu}, \bibinfo{person}{Brian Tse}, \bibinfo{person}{Jie
  Fu}, \bibinfo{person}{Stephen McAleer}, \bibinfo{person}{Yaodong Yang},
  \bibinfo{person}{Yizhou Wang}, \bibinfo{person}{Song-Chun Zhu},
  \bibinfo{person}{Yike Guo}, {and} \bibinfo{person}{Wen Gao}.}
  \bibinfo{year}{2023}\natexlab{}.
\newblock \bibinfo{title}{{AI} Alignment: A Comprehensive Survey}.
\newblock \bibinfo{howpublished}{arXiv:2310.19852}.
\newblock


\bibitem[\protect\citeauthoryear{Jobin, Ienca, and Vayena}{Jobin
  et~al\mbox{.}}{2019}]%
        {jobin2019global}
\bibfield{author}{\bibinfo{person}{Anna Jobin}, \bibinfo{person}{Marcello
  Ienca}, {and} \bibinfo{person}{Effy Vayena}.}
  \bibinfo{year}{2019}\natexlab{}.
\newblock \showarticletitle{The Global Landscape of {AI} Ethics Guidelines}.
\newblock \bibinfo{journal}{\emph{Nature Machine Intelligence}}
  \bibinfo{volume}{1}, \bibinfo{number}{9} (\bibinfo{date}{Sept.}
  \bibinfo{year}{2019}), \bibinfo{pages}{389--399}.
\newblock


\bibitem[\protect\citeauthoryear{Kandasamy, Dasarathy, Poczos, and
  Schneider}{Kandasamy et~al\mbox{.}}{2016}]%
        {kandasamy2016multi}
\bibfield{author}{\bibinfo{person}{Kirthevasan Kandasamy},
  \bibinfo{person}{Gautam Dasarathy}, \bibinfo{person}{Barnabas Poczos}, {and}
  \bibinfo{person}{Jeff Schneider}.} \bibinfo{year}{2016}\natexlab{}.
\newblock \showarticletitle{The Multi-Fidelity Multi-Armed Bandit}.
\newblock In \bibinfo{booktitle}{\emph{Advances in Neural Information
  Processing Systems}}.
\newblock


\bibitem[\protect\citeauthoryear{Kantrowitz}{Kantrowitz}{2023}]%
        {Kantrowitz2023}
\bibfield{author}{\bibinfo{person}{Alex Kantrowitz}.}
  \bibinfo{year}{2023}\natexlab{}.
\newblock \showarticletitle{The Horrific Content a {K}enyan Worker Had to See
  While Training {ChatGPT}}.
\newblock \bibinfo{journal}{\emph{Slate}} (\bibinfo{date}{May}
  \bibinfo{year}{2023}).
\newblock


\bibitem[\protect\citeauthoryear{Kirk, Bean, Vidgen, Rottger, and Hale}{Kirk
  et~al\mbox{.}}{2023a}]%
        {kirk-etal-2023-past}
\bibfield{author}{\bibinfo{person}{Hannah~Rose Kirk},
  \bibinfo{person}{Andrew~M. Bean}, \bibinfo{person}{Bertie Vidgen},
  \bibinfo{person}{Paul Rottger}, {and} \bibinfo{person}{Scott~A. Hale}.}
  \bibinfo{year}{2023}\natexlab{a}.
\newblock \showarticletitle{The Past, Present and Better Future of Feedback
  Learning in Large Language Models for Subjective Human Preferences and
  Values}. In \bibinfo{booktitle}{\emph{Proceedings of the Conference on
  Empirical Methods in Natural Language Processing}}.
  \bibinfo{pages}{2409--2430}.
\newblock


\bibitem[\protect\citeauthoryear{Kirk, Vidgen, R{\"o}ttger, and Hale}{Kirk
  et~al\mbox{.}}{2023b}]%
        {kirk2023empty}
\bibfield{author}{\bibinfo{person}{Hannah~Rose Kirk}, \bibinfo{person}{Bertie
  Vidgen}, \bibinfo{person}{Paul R{\"o}ttger}, {and} \bibinfo{person}{Scott~A.
  Hale}.} \bibinfo{year}{2023}\natexlab{b}.
\newblock \bibinfo{title}{The Empty Signifier Problem: Towards Clearer
  Paradigms for Operationalising ``Alignment''' in Large Language Models}.
\newblock \bibinfo{howpublished}{arXiv:2310.02457}.
\newblock


\bibitem[\protect\citeauthoryear{Kirk, Vidgen, R{\"o}ttger, and Hale}{Kirk
  et~al\mbox{.}}{2023c}]%
        {KirkVRH2023}
\bibfield{author}{\bibinfo{person}{Hannah~Rose Kirk}, \bibinfo{person}{Bertie
  Vidgen}, \bibinfo{person}{Paul R{\"o}ttger}, {and} \bibinfo{person}{Scott~A.
  Hale}.} \bibinfo{year}{2023}\natexlab{c}.
\newblock \bibinfo{title}{Personalisation within Bounds: A Risk Taxonomy and
  Policy Framework for the Alignment of Large Language Models with Personalised
  Feedback}.
\newblock \bibinfo{howpublished}{arXiv:2303.05453}.
\newblock


\bibitem[\protect\citeauthoryear{Klein}{Klein}{1990}]%
        {klein1990snares}
\bibfield{author}{\bibinfo{person}{Cecelia~F. Klein}.}
  \bibinfo{year}{1990}\natexlab{}.
\newblock \showarticletitle{Snares and Entrails: {M}esoamerican Symbols of Sin
  and Punishment}.
\newblock \bibinfo{journal}{\emph{Res: Anthropology and Aesthetics}}
  \bibinfo{volume}{19--20} (\bibinfo{year}{1990}), \bibinfo{pages}{81--103}.
\newblock


\bibitem[\protect\citeauthoryear{Knowles, Fledderjohann, Richards, and
  Varshney}{Knowles et~al\mbox{.}}{2023}]%
        {knowles2023trustworthy}
\bibfield{author}{\bibinfo{person}{Bran Knowles}, \bibinfo{person}{Jasmine
  Fledderjohann}, \bibinfo{person}{John~T. Richards}, {and}
  \bibinfo{person}{Kush~R. Varshney}.} \bibinfo{year}{2023}\natexlab{}.
\newblock \showarticletitle{Trustworthy {AI} and the Logics of Intersectional
  Resistance}. In \bibinfo{booktitle}{\emph{Proceedings of the ACM Conference
  on Fairness, Accountability, and Transparency}}. \bibinfo{pages}{172--182}.
\newblock


\bibitem[\protect\citeauthoryear{Kour, Zalmanovici, Zwerdling, Goldbraich,
  Fandina, Anaby-Tavor, Raz, and Farchi}{Kour et~al\mbox{.}}{2023}]%
        {kour2023unveiling}
\bibfield{author}{\bibinfo{person}{George Kour}, \bibinfo{person}{Marcel
  Zalmanovici}, \bibinfo{person}{Naama Zwerdling}, \bibinfo{person}{Esther
  Goldbraich}, \bibinfo{person}{Ora~Nova Fandina}, \bibinfo{person}{Ateret
  Anaby-Tavor}, \bibinfo{person}{Orna Raz}, {and} \bibinfo{person}{Eitan
  Farchi}.} \bibinfo{year}{2023}\natexlab{}.
\newblock \showarticletitle{Unveiling Safety Vulnerabilities of Large Language
  Models}. In \bibinfo{booktitle}{\emph{EMNLP Workshop on Generation,
  Evaluation \& Metrics}}.
\newblock


\bibitem[\protect\citeauthoryear{Krishnan, Abdilla, Moon, Souza, Adamson, Lach,
  Ghazal, Fjeld, Taylor, Havens, Jayaram, Morrow, Rizk, Ricaurte~Quijano,
  Çetin, Chatila, Dotan, Mhlambi, Jordan, and Rosenstock}{Krishnan
  et~al\mbox{.}}{[n.d.]}]%
        {Manyfesto}
\bibfield{author}{\bibinfo{person}{Aarathi Krishnan}, \bibinfo{person}{Angie
  Abdilla}, \bibinfo{person}{A~Jung Moon}, \bibinfo{person}{Carlos~Affonso
  Souza}, \bibinfo{person}{Chelle Adamson}, \bibinfo{person}{Eileen~M. Lach},
  \bibinfo{person}{Farah Ghazal}, \bibinfo{person}{Jessica Fjeld},
  \bibinfo{person}{Jennyfer Taylor}, \bibinfo{person}{John~C. Havens},
  \bibinfo{person}{Malavika Jayaram}, \bibinfo{person}{Monique Morrow},
  \bibinfo{person}{Nagla Rizk}, \bibinfo{person}{Paola Ricaurte~Quijano},
  \bibinfo{person}{R.~Buse Çetin}, \bibinfo{person}{Raja Chatila},
  \bibinfo{person}{Ravit Dotan}, \bibinfo{person}{Sabelo Mhlambi},
  \bibinfo{person}{Sara Jordan}, {and} \bibinfo{person}{Sarita Rosenstock}.}
  \bibinfo{year}{[n.d.]}\natexlab{}.
\newblock \bibinfo{title}{{AI} Decolonial Manyfesto}.
\newblock \bibinfo{howpublished}{https://manyfesto.ai/}.
\newblock


\bibitem[\protect\citeauthoryear{Lambright}{Lambright}{2019}]%
        {lambright2019digital}
\bibfield{author}{\bibinfo{person}{Katie Lambright}.}
  \bibinfo{year}{2019}\natexlab{}.
\newblock \showarticletitle{Digital Redlining: The {N}extdoor App and the
  Neighborhood of Make-Believe}.
\newblock \bibinfo{journal}{\emph{Cultural Critique}}  \bibinfo{volume}{103}
  (\bibinfo{date}{Spring} \bibinfo{year}{2019}), \bibinfo{pages}{84--90}.
\newblock


\bibitem[\protect\citeauthoryear{Lera-Leri, Bistaffa, Serramia, Lopez-Sanchez,
  and Rodriguez-Aguilar}{Lera-Leri et~al\mbox{.}}{2022}]%
        {lera2022towards}
\bibfield{author}{\bibinfo{person}{Roger Lera-Leri}, \bibinfo{person}{Filippo
  Bistaffa}, \bibinfo{person}{Marc Serramia}, \bibinfo{person}{Maite
  Lopez-Sanchez}, {and} \bibinfo{person}{Juan Rodriguez-Aguilar}.}
  \bibinfo{year}{2022}\natexlab{}.
\newblock \showarticletitle{Towards Pluralistic Value Alignment: Aggregating
  Value Systems Through $\ell_p$-Regression}. In
  \bibinfo{booktitle}{\emph{Proceedings of the International Conference on
  Autonomous Agents and Multiagent Systems}}. \bibinfo{pages}{780--788}.
\newblock


\bibitem[\protect\citeauthoryear{Liao and Wortman~Vaughan}{Liao and
  Wortman~Vaughan}{2023}]%
        {liao2023ai}
\bibfield{author}{\bibinfo{person}{Q.~Vera Liao} {and}
  \bibinfo{person}{Jennifer Wortman~Vaughan}.} \bibinfo{year}{2023}\natexlab{}.
\newblock \bibinfo{title}{{AI} Transparency in the Age of {LLMs}: A
  Human-Centered Research Roadmap}.
\newblock \bibinfo{howpublished}{arXiv:2306.01941}.
\newblock


\bibitem[\protect\citeauthoryear{Lin}{Lin}{2023}]%
        {lin2023all}
\bibfield{author}{\bibinfo{person}{Chien-Te Lin}.}
  \bibinfo{year}{2023}\natexlab{}.
\newblock \showarticletitle{All About the Human: A {B}uddhist Take on {AI}
  Ethics}.
\newblock \bibinfo{journal}{\emph{Business Ethics, the Environment \&
  Responsibility}} \bibinfo{volume}{32}, \bibinfo{number}{3}
  (\bibinfo{date}{July} \bibinfo{year}{2023}), \bibinfo{pages}{1113--1122}.
\newblock


\bibitem[\protect\citeauthoryear{Lindgren}{Lindgren}{2023}]%
        {lindgren2023handbook}
\bibfield{editor}{\bibinfo{person}{Simon Lindgren}} (Ed.).
  \bibinfo{year}{2023}\natexlab{}.
\newblock \bibinfo{booktitle}{\emph{Handbook of Critical Studies of Artificial
  Intelligence}}.
\newblock \bibinfo{publisher}{Edward Elgar Publishing},
  \bibinfo{address}{Cheltenham, UK}.
\newblock


\bibitem[\protect\citeauthoryear{Maheshwarachary}{Maheshwarachary}{1988}]%
        {Maheshwarachary1988}
\bibfield{author}{\bibinfo{person}{Maheshwarachary}.}
  \bibinfo{year}{1988}\natexlab{}.
\newblock \bibinfo{booktitle}{\emph{What Have We Learnt?}}
\newblock \bibinfo{publisher}{Nav Yug Press}, \bibinfo{address}{Aligarh, UP,
  India}.
\newblock


\bibitem[\protect\citeauthoryear{Maitra}{Maitra}{2020}]%
        {maitra2020artificial}
\bibfield{author}{\bibinfo{person}{Suvradip Maitra}.}
  \bibinfo{year}{2020}\natexlab{}.
\newblock \showarticletitle{Artificial Intelligence and Indigenous
  Perspectives: Protecting and Empowering Intelligent Human Beings}. In
  \bibinfo{booktitle}{\emph{Proceedings of the AAAI/ACM Conference on AI,
  Ethics, and Society}}. \bibinfo{pages}{320--326}.
\newblock


\bibitem[\protect\citeauthoryear{Maldonado-Torres}{Maldonado-Torres}{2007}]%
        {maldonado2007coloniality}
\bibfield{author}{\bibinfo{person}{Nelson Maldonado-Torres}.}
  \bibinfo{year}{2007}\natexlab{}.
\newblock \showarticletitle{On the Coloniality of Being: Contributions to the
  Development of a Concept}.
\newblock \bibinfo{journal}{\emph{Cultural Studies}} \bibinfo{volume}{21},
  \bibinfo{number}{2-3} (\bibinfo{date}{April} \bibinfo{year}{2007}),
  \bibinfo{pages}{240--270}.
\newblock


\bibitem[\protect\citeauthoryear{Maldonado-Torres}{Maldonado-Torres}{2017}]%
        {maldonado2017religion}
\bibfield{author}{\bibinfo{person}{Nelson Maldonado-Torres}.}
  \bibinfo{year}{2017}\natexlab{}.
\newblock \showarticletitle{Religion, Modernity, and Coloniality}.
\newblock In \bibinfo{booktitle}{\emph{Religion, Theory, Critique: Classic and
  Contemporary Approaches and Methodologies}},
  \bibfield{editor}{\bibinfo{person}{Richard King}} (Ed.).
  \bibinfo{publisher}{Columbia University Press}, \bibinfo{address}{New York,
  NY, USA}, \bibinfo{pages}{547--554}.
\newblock


\bibitem[\protect\citeauthoryear{Maldonado-Torres}{Maldonado-Torres}{2021}]%
        {maldonado2021coloniality}
\bibfield{author}{\bibinfo{person}{Nelson Maldonado-Torres}.}
  \bibinfo{year}{2021}\natexlab{}.
\newblock \showarticletitle{On the Coloniality of Human Rights}.
\newblock In \bibinfo{booktitle}{\emph{The Pluriverse of Human Rights: The
  Diversity of Struggles for Dignity}},
  \bibfield{editor}{\bibinfo{person}{Boaventura De~Sousa~Santos} {and}
  \bibinfo{person}{Bruno Martins}} (Eds.). \bibinfo{publisher}{Routledge},
  \bibinfo{address}{New York, NY, USA}, \bibinfo{pages}{62--82}.
\newblock


\bibitem[\protect\citeauthoryear{Martin, Prabhakaran, Kuhlberg, Smart, and
  Isaac}{Martin et~al\mbox{.}}{2020}]%
        {martin2020extending}
\bibfield{author}{\bibinfo{person}{Donald Martin, Jr.},
  \bibinfo{person}{Vinodkumar Prabhakaran}, \bibinfo{person}{Jill Kuhlberg},
  \bibinfo{person}{Andrew Smart}, {and} \bibinfo{person}{William~S. Isaac}.}
  \bibinfo{year}{2020}\natexlab{}.
\newblock \bibinfo{title}{Extending the Machine Learning Abstraction Boundary:
  A Complex Systems Approach to Incorporate Societal Context}.
\newblock \bibinfo{howpublished}{arXiv:2006.09663}.
\newblock


\bibitem[\protect\citeauthoryear{Marx}{Marx}{[n.d.]}]%
        {marx2023economic}
\bibfield{author}{\bibinfo{person}{Karl Marx}.}
  \bibinfo{year}{[n.d.]}\natexlab{}.
\newblock \bibinfo{booktitle}{\emph{Economic and Philosophic Manuscripts of
  1844}}.
\newblock


\bibitem[\protect\citeauthoryear{Mastin}{Mastin}{2009}]%
        {Mastin2009}
\bibfield{author}{\bibinfo{person}{Luke Mastin}.}
  \bibinfo{year}{2009}\natexlab{}.
\newblock \bibinfo{title}{Moral Universalism}.
\newblock
  \bibinfo{howpublished}{https://www.philosophybasics.com/branch\_moral\_universalism.html}.
\newblock


\bibitem[\protect\citeauthoryear{Mehrotra}{Mehrotra}{2022}]%
        {Mehrotra2022}
\bibfield{author}{\bibinfo{person}{Karishma Mehrotra}.}
  \bibinfo{year}{2022}\natexlab{}.
\newblock \showarticletitle{Human Touch}.
\newblock \bibinfo{journal}{\emph{Fifty Two}} (\bibinfo{date}{July}
  \bibinfo{year}{2022}).
\newblock


\bibitem[\protect\citeauthoryear{Menon}{Menon}{2022}]%
        {menon2022debunking}
\bibfield{author}{\bibinfo{person}{Annapurna Menon}.}
  \bibinfo{year}{2022}\natexlab{}.
\newblock \showarticletitle{Debunking {H}indutva Appropriation of Decolonial
  Thought}.
\newblock \bibinfo{journal}{\emph{Interfere: Journal for Critical Thought and
  Radical Politics}}  \bibinfo{volume}{3} (\bibinfo{year}{2022}),
  \bibinfo{pages}{36--57}.
\newblock


\bibitem[\protect\citeauthoryear{Mhlambi and Tiribelli}{Mhlambi and
  Tiribelli}{2023}]%
        {mhlambi2023decolonizing}
\bibfield{author}{\bibinfo{person}{S\'{a}b{\"{e}}lo Mhlambi} {and}
  \bibinfo{person}{Simona Tiribelli}.} \bibinfo{year}{2023}\natexlab{}.
\newblock \showarticletitle{Decolonizing {AI} Ethics: Relational Autonomy as a
  Means to Counter {AI} Harms}.
\newblock \bibinfo{journal}{\emph{Topoi}} \bibinfo{volume}{42},
  \bibinfo{number}{3} (\bibinfo{date}{July} \bibinfo{year}{2023}),
  \bibinfo{pages}{867--880}.
\newblock


\bibitem[\protect\citeauthoryear{Miceli and Posada}{Miceli and Posada}{2022}]%
        {miceli2022data}
\bibfield{author}{\bibinfo{person}{Milagros Miceli} {and}
  \bibinfo{person}{Julian Posada}.} \bibinfo{year}{2022}\natexlab{}.
\newblock \showarticletitle{The Data-Production Dispositif}. In
  \bibinfo{booktitle}{\emph{Proceedings of the ACM on Human-Computer
  Interaction CSCW2}}. \bibinfo{pages}{460}.
\newblock


\bibitem[\protect\citeauthoryear{Mignolo}{Mignolo}{2010}]%
        {mignolo2010introduction}
\bibfield{author}{\bibinfo{person}{Walter~D. Mignolo}.}
  \bibinfo{year}{2010}\natexlab{}.
\newblock \showarticletitle{Introduction: Coloniality of Power and De-Colonial
  Thinking}.
\newblock In \bibinfo{booktitle}{\emph{Globalization and the Decolonial
  Option}}, \bibfield{editor}{\bibinfo{person}{Walter~D. Mignolo} {and}
  \bibinfo{person}{Arturo Escobar}} (Eds.). \bibinfo{publisher}{Routledge},
  \bibinfo{address}{London, UK}, \bibinfo{pages}{1--21}.
\newblock


\bibitem[\protect\citeauthoryear{Mignolo and Walsh}{Mignolo and Walsh}{2018}]%
        {mignolo2018decoloniality}
\bibfield{author}{\bibinfo{person}{Walter~D. Mignolo} {and}
  \bibinfo{person}{Catherine~E. Walsh}.} \bibinfo{year}{2018}\natexlab{}.
\newblock \bibinfo{booktitle}{\emph{On Decoloniality: Concepts, Analytics,
  Praxis}}.
\newblock \bibinfo{publisher}{Duke University Press}, \bibinfo{address}{Durham,
  NC, USA}.
\newblock


\bibitem[\protect\citeauthoryear{Mitova}{Mitova}{2021}]%
        {mitova2021decolonise}
\bibfield{author}{\bibinfo{person}{Veli Mitova}.}
  \bibinfo{year}{2021}\natexlab{}.
\newblock \showarticletitle{How to Decolonise Knowledge Without Too Much
  Relativism}.
\newblock In \bibinfo{booktitle}{\emph{Decolonisation as Democratisation:
  Global Insights into the {S}outh {A}frican Experience}},
  \bibfield{editor}{\bibinfo{person}{Siseko~H. Kumalo}} (Ed.).
  \bibinfo{publisher}{Lynne Rienner Publishers}.
\newblock


\bibitem[\protect\citeauthoryear{Mohamed, Png, and Isaac}{Mohamed
  et~al\mbox{.}}{2020}]%
        {mohamed2020decolonial}
\bibfield{author}{\bibinfo{person}{Shakir Mohamed},
  \bibinfo{person}{Marie-Therese Png}, {and} \bibinfo{person}{William Isaac}.}
  \bibinfo{year}{2020}\natexlab{}.
\newblock \showarticletitle{Decolonial {AI}: Decolonial Theory as
  Sociotechnical Foresight in Artificial Intelligence}.
\newblock \bibinfo{journal}{\emph{Philosophy \& Technology}}
  \bibinfo{volume}{33} (\bibinfo{date}{July} \bibinfo{year}{2020}),
  \bibinfo{pages}{659--684}.
\newblock


\bibitem[\protect\citeauthoryear{M{\"o}kander, Schuett, Kirk, and
  Floridi}{M{\"o}kander et~al\mbox{.}}{2023}]%
        {mokander2023auditing}
\bibfield{author}{\bibinfo{person}{Jakob M{\"o}kander}, \bibinfo{person}{Jonas
  Schuett}, \bibinfo{person}{Hannah~Rose Kirk}, {and} \bibinfo{person}{Luciano
  Floridi}.} \bibinfo{year}{2023}\natexlab{}.
\newblock \showarticletitle{Auditing Large Language Models: A Three-Layered
  Approach}.
\newblock \bibinfo{journal}{\emph{AI and Ethics}} (\bibinfo{year}{2023}).
\newblock


\bibitem[\protect\citeauthoryear{M{\"o}ller}{M{\"o}ller}{2016}]%
        {Moller2016}
\bibfield{author}{\bibinfo{person}{Niklas M{\"o}ller}.}
  \bibinfo{year}{2016}\natexlab{}.
\newblock \showarticletitle{Value Uncertainty}.
\newblock In \bibinfo{booktitle}{\emph{The Argumentative Turn in Policy
  Analysis: Reasoning About Uncertainty}},
  \bibfield{editor}{\bibinfo{person}{Sven~Ove Hansson} {and}
  \bibinfo{person}{Gertrude~Hirsch Hadorn}} (Eds.).
  \bibinfo{publisher}{Springer}, \bibinfo{address}{Switzerland},
  \bibinfo{pages}{105--133}.
\newblock


\bibitem[\protect\citeauthoryear{Mondal}{Mondal}{2012}]%
        {Mondal2012}
\bibfield{author}{\bibinfo{person}{Parthasarathi Mondal}.}
  \bibinfo{year}{2012}\natexlab{}.
\newblock \showarticletitle{Philosophy and Nationalism in {I}ndia: A
  Preliminary Essay}.
\newblock \bibinfo{journal}{\emph{Journal of Social Work and Social
  Development}} \bibinfo{volume}{3}, \bibinfo{number}{1--2}
  (\bibinfo{date}{June--Dec.} \bibinfo{year}{2012}).
\newblock


\bibitem[\protect\citeauthoryear{Motilal, Maitra, and Prajapati}{Motilal
  et~al\mbox{.}}{2021}]%
        {motilal2021care}
\bibfield{author}{\bibinfo{person}{Shashi Motilal}, \bibinfo{person}{Keya
  Maitra}, {and} \bibinfo{person}{Prakriti Prajapati}.}
  \bibinfo{year}{2021}\natexlab{}.
\newblock \showarticletitle{Care, Community, Compassion and Virtue:
  Decolonizing Our Moral Landscape}.
\newblock In \bibinfo{booktitle}{\emph{The Ethics of Governance: Moral Limits
  of Policy Decisions}}. \bibinfo{publisher}{Springer},
  \bibinfo{address}{Singapore}, \bibinfo{pages}{141--176}.
\newblock


\bibitem[\protect\citeauthoryear{Muir}{Muir}{2021}]%
        {muir2021hci}
\bibfield{author}{\bibinfo{person}{Alexander Muir}.}
  \bibinfo{year}{2021}\natexlab{}.
\newblock \showarticletitle{Where {HCI} Meets the Spiritual Path: The Three
  Yogas of the Bhagavad G{\=\i}t{\=a}}. In \bibinfo{booktitle}{\emph{Extended
  Abstracts of the CHI Conference on Human Factors in Computing Systems}}.
  \bibinfo{pages}{38}.
\newblock


\bibitem[\protect\citeauthoryear{Muldoon and Wu}{Muldoon and Wu}{2023}]%
        {muldoon2023artificial}
\bibfield{author}{\bibinfo{person}{James Muldoon} {and}
  \bibinfo{person}{Boxi~A. Wu}.} \bibinfo{year}{2023}\natexlab{}.
\newblock \showarticletitle{Artificial Intelligence in the Colonial Matrix of
  Power}.
\newblock \bibinfo{journal}{\emph{Philosophy \& Technology}}
  \bibinfo{volume}{36}, \bibinfo{number}{4} (\bibinfo{date}{Dec.}
  \bibinfo{year}{2023}), \bibinfo{pages}{80}.
\newblock


\bibitem[\protect\citeauthoryear{Nagireddy, Chiazor, Singh, and
  Baldini}{Nagireddy et~al\mbox{.}}{2024}]%
        {nagireddy2024socialstigmaqa}
\bibfield{author}{\bibinfo{person}{Manish Nagireddy}, \bibinfo{person}{Lamogha
  Chiazor}, \bibinfo{person}{Moninder Singh}, {and} \bibinfo{person}{Ioana
  Baldini}.} \bibinfo{year}{2024}\natexlab{}.
\newblock \showarticletitle{{SocialStigmaQA}: A Benchmark to Uncover Stigma
  Amplification in Generative Language Models}. In
  \bibinfo{booktitle}{\emph{Proceedings of the AAAI Conference on Artificial
  Intelligence}}, Vol.~\bibinfo{volume}{38}. \bibinfo{pages}{21454--21462}.
\newblock


\bibitem[\protect\citeauthoryear{Na'puti and Cruz}{Na'puti and Cruz}{2022}]%
        {na2022mapping}
\bibfield{author}{\bibinfo{person}{Tiara~R Na'puti} {and}
  \bibinfo{person}{Joelle~M. Cruz}.} \bibinfo{year}{2022}\natexlab{}.
\newblock \showarticletitle{Mapping Interventions: Toward a Decolonial and
  Indigenous Praxis Across Communication Subfields}.
\newblock \bibinfo{journal}{\emph{Communication, Culture and Critique}}
  \bibinfo{volume}{15}, \bibinfo{number}{1} (\bibinfo{date}{March}
  \bibinfo{year}{2022}), \bibinfo{pages}{1--20}.
\newblock


\bibitem[\protect\citeauthoryear{Nix}{Nix}{2023}]%
        {Nix2023}
\bibfield{author}{\bibinfo{person}{Jessica Nix}.}
  \bibinfo{year}{2023}\natexlab{}.
\newblock \showarticletitle{One {AI} Startup Wants to Tackle Bias by Teaching
  Black History: Equality}.
\newblock
  \bibinfo{howpublished}{https://www.bloomberg.com/news/newsletters/2023-10-19/latimer-ai-startup-for-hcbus-trains-model-on-african-american-history-culture}.
\newblock \bibinfo{journal}{\emph{Bloomberg Newsletter}} (\bibinfo{date}{Oct.}
  \bibinfo{year}{2023}).
\newblock


\bibitem[\protect\citeauthoryear{Noothigattu, Bouneffouf, Mattei, Chandra,
  Madan, Varshney, Campbell, Singh, and Rossi}{Noothigattu
  et~al\mbox{.}}{2019}]%
        {noothigattu2019teaching}
\bibfield{author}{\bibinfo{person}{Ritesh Noothigattu},
  \bibinfo{person}{Djallel Bouneffouf}, \bibinfo{person}{Nicholas Mattei},
  \bibinfo{person}{Rachita Chandra}, \bibinfo{person}{Piyush Madan},
  \bibinfo{person}{Kush~R. Varshney}, \bibinfo{person}{Murray Campbell},
  \bibinfo{person}{Moninder Singh}, {and} \bibinfo{person}{Francesca Rossi}.}
  \bibinfo{year}{2019}\natexlab{}.
\newblock \showarticletitle{Teaching {AI} Agents Ethical Values Using
  Reinforcement Learning and Policy Orchestration}.
\newblock \bibinfo{journal}{\emph{IBM Journal of Research and Development}}
  \bibinfo{volume}{63}, \bibinfo{number}{4/5} (\bibinfo{date}{July--Sept.}
  \bibinfo{year}{2019}), \bibinfo{pages}{2}.
\newblock


\bibitem[\protect\citeauthoryear{Nwankwo and Sonna}{Nwankwo and Sonna}{2019}]%
        {nwankwo2019africa}
\bibfield{author}{\bibinfo{person}{Ezinne Nwankwo} {and}
  \bibinfo{person}{Belona Sonna}.} \bibinfo{year}{2019}\natexlab{}.
\newblock \showarticletitle{{A}frica's Social Contract with {AI}}.
\newblock \bibinfo{journal}{\emph{ACM XRDS: Crossroads}} \bibinfo{volume}{26},
  \bibinfo{number}{2} (\bibinfo{date}{Winter} \bibinfo{year}{2019}),
  \bibinfo{pages}{44--48}.
\newblock


\bibitem[\protect\citeauthoryear{Ouyang, Wu, Jiang, Almeida, Wainwright,
  Mishkin, Zhang, Agarwal, Slama, Ray, Schulman, Hilton, Kelton, Miller,
  Simens, Askell, Welinder, Christiano, Leike, and Lowe}{Ouyang
  et~al\mbox{.}}{2022}]%
        {Ouyang2022}
\bibfield{author}{\bibinfo{person}{Long Ouyang}, \bibinfo{person}{Jeff Wu},
  \bibinfo{person}{Xu Jiang}, \bibinfo{person}{Diogo Almeida},
  \bibinfo{person}{Carroll~L. Wainwright}, \bibinfo{person}{Pamela Mishkin},
  \bibinfo{person}{Chong Zhang}, \bibinfo{person}{Sandhini Agarwal},
  \bibinfo{person}{Katarina Slama}, \bibinfo{person}{Alex Ray},
  \bibinfo{person}{John Schulman}, \bibinfo{person}{Jacob Hilton},
  \bibinfo{person}{Fraser Kelton}, \bibinfo{person}{Luke Miller},
  \bibinfo{person}{Maddie Simens}, \bibinfo{person}{Amanda Askell},
  \bibinfo{person}{Peter Welinder}, \bibinfo{person}{Paul Christiano},
  \bibinfo{person}{Jan Leike}, {and} \bibinfo{person}{Ryan Lowe}.}
  \bibinfo{year}{2022}\natexlab{}.
\newblock \showarticletitle{Training Language Models to Follow instructions
  with Human Feedback}. In \bibinfo{booktitle}{\emph{Advances in Neural
  Information Processing Systems}}. \bibinfo{pages}{27730--27744}.
\newblock


\bibitem[\protect\citeauthoryear{Patel}{Patel}{2020}]%
        {patel2020race}
\bibfield{author}{\bibinfo{person}{Kamna Patel}.}
  \bibinfo{year}{2020}\natexlab{}.
\newblock \showarticletitle{Race and a Decolonial Turn in Development Studies}.
\newblock \bibinfo{journal}{\emph{Third World Quarterly}} \bibinfo{volume}{41},
  \bibinfo{number}{9} (\bibinfo{date}{July} \bibinfo{year}{2020}),
  \bibinfo{pages}{1463--1475}.
\newblock


\bibitem[\protect\citeauthoryear{Patin, Sebastian, Yeon, Bertolini, and
  Grimm}{Patin et~al\mbox{.}}{2021}]%
        {patin2021interrupting}
\bibfield{author}{\bibinfo{person}{Beth Patin}, \bibinfo{person}{Melinda
  Sebastian}, \bibinfo{person}{Jieun Yeon}, \bibinfo{person}{Danielle
  Bertolini}, {and} \bibinfo{person}{Alexandra Grimm}.}
  \bibinfo{year}{2021}\natexlab{}.
\newblock \showarticletitle{Interrupting Epistemicide: A Practical Framework
  for Naming, Identifying, and Ending Epistemic Injustice in the Information
  Professions}.
\newblock \bibinfo{journal}{\emph{Journal of the Association for Information
  Science and Technology}} \bibinfo{volume}{72}, \bibinfo{number}{10}
  (\bibinfo{date}{Oct.} \bibinfo{year}{2021}), \bibinfo{pages}{1306--1318}.
\newblock


\bibitem[\protect\citeauthoryear{Pendse, Nkemelu, Bidwell, Jadhav, Pathare,
  De~Choudhury, and Kumar}{Pendse et~al\mbox{.}}{2022}]%
        {pendse2022treatment}
\bibfield{author}{\bibinfo{person}{Sachin~R. Pendse}, \bibinfo{person}{Daniel
  Nkemelu}, \bibinfo{person}{Nicola~J. Bidwell}, \bibinfo{person}{Sushrut
  Jadhav}, \bibinfo{person}{Soumitra Pathare}, \bibinfo{person}{Munmun
  De~Choudhury}, {and} \bibinfo{person}{Neha Kumar}.}
  \bibinfo{year}{2022}\natexlab{}.
\newblock \showarticletitle{From Treatment to Healing: Envisioning a Decolonial
  Digital Mental Health}. In \bibinfo{booktitle}{\emph{Proceedings of the CHI
  Conference on Human Factors in Computing Systems}}. \bibinfo{pages}{548}.
\newblock


\bibitem[\protect\citeauthoryear{Perrigo}{Perrigo}{2023a}]%
        {Perrigo2023}
\bibfield{author}{\bibinfo{person}{Billy Perrigo}.}
  \bibinfo{year}{2023}\natexlab{a}.
\newblock \showarticletitle{{OpenAI} Used {K}enyan Workers on Less Than \$2 Per
  Hour to Make {ChatGPT} Less Toxic}.
\newblock \bibinfo{journal}{\emph{Time}} (\bibinfo{date}{Jan.}
  \bibinfo{year}{2023}).
\newblock


\bibitem[\protect\citeauthoryear{Perrigo}{Perrigo}{2023b}]%
        {Perrigo2023b}
\bibfield{author}{\bibinfo{person}{Billy Perrigo}.}
  \bibinfo{year}{2023}\natexlab{b}.
\newblock \showarticletitle{The Workers Behind {AI} Rarely See Its Rewards.
  {T}his {I}ndian Startup Wants to Fix That}.
\newblock \bibinfo{journal}{\emph{Time}} (\bibinfo{date}{July}
  \bibinfo{year}{2023}).
\newblock


\bibitem[\protect\citeauthoryear{Press}{Press}{2023}]%
        {alliance2023}
\bibfield{author}{\bibinfo{person}{Associated Press}.}
  \bibinfo{year}{2023}\natexlab{}.
\newblock \showarticletitle{Meta and {IBM} Launch ‘{AI Alliance}’ to
  Promote Open-Source {AI} Development}.
\newblock \bibinfo{journal}{\emph{The Guardian}} (\bibinfo{date}{Dec.}
  \bibinfo{year}{2023}).
\newblock


\bibitem[\protect\citeauthoryear{Qi, Zeng, Xie, Chen, Jia, Mittal, and
  Henderson}{Qi et~al\mbox{.}}{2023}]%
        {qi2023fine}
\bibfield{author}{\bibinfo{person}{Xiangyu Qi}, \bibinfo{person}{Yi Zeng},
  \bibinfo{person}{Tinghao Xie}, \bibinfo{person}{Pin-Yu Chen},
  \bibinfo{person}{Ruoxi Jia}, \bibinfo{person}{Prateek Mittal}, {and}
  \bibinfo{person}{Peter Henderson}.} \bibinfo{year}{2023}\natexlab{}.
\newblock \bibinfo{title}{Fine-Tuning Aligned Language Models Compromises
  Safety, Even When Users Do Not Intend To!}
\newblock \bibinfo{howpublished}{arXiv:2310.03693}.
\newblock


\bibitem[\protect\citeauthoryear{Quijano}{Quijano}{2007}]%
        {quijano2007coloniality}
\bibfield{author}{\bibinfo{person}{An{\'\i}bal Quijano}.}
  \bibinfo{year}{2007}\natexlab{}.
\newblock \showarticletitle{Coloniality and Modernity/Rationality}.
\newblock \bibinfo{journal}{\emph{Cultural Studies}} \bibinfo{volume}{21},
  \bibinfo{number}{2-3} (\bibinfo{date}{April} \bibinfo{year}{2007}),
  \bibinfo{pages}{168--178}.
\newblock


\bibitem[\protect\citeauthoryear{Raji, Bender, Paullada, Denton, and
  Hanna}{Raji et~al\mbox{.}}{2021}]%
        {raji2021ai}
\bibfield{author}{\bibinfo{person}{Inioluwa~Deborah Raji},
  \bibinfo{person}{Emily~M. Bender}, \bibinfo{person}{Amandalynne Paullada},
  \bibinfo{person}{Emily Denton}, {and} \bibinfo{person}{Alex Hanna}.}
  \bibinfo{year}{2021}\natexlab{}.
\newblock \showarticletitle{{AI} and the Everything in the Whole Wide World
  Benchmark}. In \bibinfo{booktitle}{\emph{Proceedings of the Neural
  Information Processing Systems Track on Datasets and Benchmarks}}.
\newblock


\bibitem[\protect\citeauthoryear{Ranganathan}{Ranganathan}{2016}]%
        {Ranganathan2016-RANNAM-3}
\bibfield{author}{\bibinfo{person}{Shyam Ranganathan}.}
  \bibinfo{year}{2016}\natexlab{}.
\newblock \showarticletitle{N\={a}g\={a}rjuna and Madhy\={a}maka Ethics}.
\newblock In \bibinfo{booktitle}{\emph{Philosophy, E-PG Pathshala}},
  \bibfield{editor}{\bibinfo{person}{A.~Raghuramaraju}} (Ed.).
  \bibinfo{publisher}{India National Mission on Education through Information
  and Communication Technology}, \bibinfo{address}{Delhi}.
\newblock


\bibitem[\protect\citeauthoryear{Ranganathan}{Ranganathan}{2022}]%
        {ranganathan2022hinduism}
\bibfield{author}{\bibinfo{person}{Shyam Ranganathan}.}
  \bibinfo{year}{2022}\natexlab{}.
\newblock \showarticletitle{Hinduism, Belief and the Colonial Invention of
  Religion: A Before and After Comparison}.
\newblock \bibinfo{journal}{\emph{Religions}} \bibinfo{volume}{13},
  \bibinfo{number}{10} (\bibinfo{date}{Sept.} \bibinfo{year}{2022}),
  \bibinfo{pages}{891}.
\newblock


\bibitem[\protect\citeauthoryear{Rawls}{Rawls}{2009}]%
        {rawls2009theory}
\bibfield{author}{\bibinfo{person}{John Rawls}.}
  \bibinfo{year}{2009}\natexlab{}.
\newblock \bibinfo{booktitle}{\emph{A Theory of Justice}}.
\newblock \bibinfo{publisher}{Harvard University Press},
  \bibinfo{address}{Cambridge, MA, USA}.
\newblock


\bibitem[\protect\citeauthoryear{Reyes}{Reyes}{2015}]%
        {reyes2015loob}
\bibfield{author}{\bibinfo{person}{Jeremiah Reyes}.}
  \bibinfo{year}{2015}\natexlab{}.
\newblock \showarticletitle{{L}o\'ob and {K}apwa: An Introduction to a
  {F}ilipino Virtue Ethics}.
\newblock \bibinfo{journal}{\emph{Asian Philosophy}} \bibinfo{volume}{25},
  \bibinfo{number}{2} (\bibinfo{date}{June} \bibinfo{year}{2015}),
  \bibinfo{pages}{148--171}.
\newblock


\bibitem[\protect\citeauthoryear{Ricaurte}{Ricaurte}{2019}]%
        {ricaurte2019data}
\bibfield{author}{\bibinfo{person}{Paola Ricaurte}.}
  \bibinfo{year}{2019}\natexlab{}.
\newblock \showarticletitle{Data Epistemologies, the Coloniality of Power, and
  Resistance}.
\newblock \bibinfo{journal}{\emph{Television \& New Media}}
  \bibinfo{volume}{20}, \bibinfo{number}{4} (\bibinfo{date}{May}
  \bibinfo{year}{2019}), \bibinfo{pages}{350--365}.
\newblock


\bibitem[\protect\citeauthoryear{Ricaurte}{Ricaurte}{2022}]%
        {ricaurte2022ethics}
\bibfield{author}{\bibinfo{person}{Paola Ricaurte}.}
  \bibinfo{year}{2022}\natexlab{}.
\newblock \showarticletitle{Ethics for the Majority World: {AI} and the
  Question of Violence at Scale}.
\newblock \bibinfo{journal}{\emph{Media, Culture \& Society}}
  \bibinfo{volume}{44}, \bibinfo{number}{4} (\bibinfo{date}{May}
  \bibinfo{year}{2022}), \bibinfo{pages}{726--745}.
\newblock


\bibitem[\protect\citeauthoryear{Schrei}{Schrei}{2010}]%
        {Schrei2010}
\bibfield{author}{\bibinfo{person}{Josh Schrei}.}
  \bibinfo{year}{2010}\natexlab{}.
\newblock \bibinfo{title}{The God Project: Hinduism as Open-Source Faith}.
\newblock
  \bibinfo{howpublished}{https://www.huffpost.com/entry/the-god-project-hinduism\_b\_486099}.
\newblock


\bibitem[\protect\citeauthoryear{Selbst, boyd, Friedler, Venkatasubramanian,
  and Vertesi}{Selbst et~al\mbox{.}}{2019}]%
        {selbst2019fairness}
\bibfield{author}{\bibinfo{person}{Andrew~D. Selbst}, \bibinfo{person}{danah
  boyd}, \bibinfo{person}{Sorelle~A. Friedler}, \bibinfo{person}{Suresh
  Venkatasubramanian}, {and} \bibinfo{person}{Janet Vertesi}.}
  \bibinfo{year}{2019}\natexlab{}.
\newblock \showarticletitle{Fairness and Abstraction in Sociotechnical
  Systems}. In \bibinfo{booktitle}{\emph{Proceedings of the Conference on
  Fairness, Accountability, and Transparency}}. \bibinfo{pages}{59--68}.
\newblock


\bibitem[\protect\citeauthoryear{Sen}{Sen}{2005}]%
        {sen2012argumentative}
\bibfield{author}{\bibinfo{person}{Amartya Sen}.}
  \bibinfo{year}{2005}\natexlab{}.
\newblock \bibinfo{booktitle}{\emph{The Argumentative Indian: Writings on
  Indian History, Culture and Identity}}.
\newblock \bibinfo{publisher}{Penguin Books}.
\newblock


\bibitem[\protect\citeauthoryear{Shani and Chadha~Behera}{Shani and
  Chadha~Behera}{2022}]%
        {shani2022provincialising}
\bibfield{author}{\bibinfo{person}{Giorgio Shani} {and}
  \bibinfo{person}{Navnita Chadha~Behera}.} \bibinfo{year}{2022}\natexlab{}.
\newblock \showarticletitle{Provincialising International Relations through a
  Reading of Dharma}.
\newblock \bibinfo{journal}{\emph{Review of International Studies}}
  \bibinfo{volume}{48}, \bibinfo{number}{5} (\bibinfo{date}{Dec.}
  \bibinfo{year}{2022}), \bibinfo{pages}{837--856}.
\newblock


\bibitem[\protect\citeauthoryear{Shelby, Rismani, Henne, Moon, Rostamzadeh,
  Nicholas, Yilla, Gallegos, Smart, Garcia, and Virk}{Shelby
  et~al\mbox{.}}{2022}]%
        {shelby2022sociotechnical}
\bibfield{author}{\bibinfo{person}{Renee Shelby}, \bibinfo{person}{Shalaleh
  Rismani}, \bibinfo{person}{Kathryn Henne}, \bibinfo{person}{AJung Moon},
  \bibinfo{person}{Negar Rostamzadeh}, \bibinfo{person}{Paul Nicholas},
  \bibinfo{person}{N'Mah Yilla}, \bibinfo{person}{Jess Gallegos},
  \bibinfo{person}{Andrew Smart}, \bibinfo{person}{Emilio Garcia}, {and}
  \bibinfo{person}{Gurleen Virk}.} \bibinfo{year}{2022}\natexlab{}.
\newblock \bibinfo{title}{Sociotechnical Harms: Scoping a Taxonomy for Harm
  Reduction}.
\newblock \bibinfo{howpublished}{arXiv:2210.05791}.
\newblock


\bibitem[\protect\citeauthoryear{Shil}{Shil}{2020}]%
        {Shil2020}
\bibfield{author}{\bibinfo{person}{Partha~Pratim Shil}.}
  \bibinfo{year}{2020}\natexlab{}.
\newblock \showarticletitle{The {I}ndian Liberal Nostalgia for a Tolerant
  {H}induism Is Misplaced}.
\newblock \bibinfo{journal}{\emph{The Wire}} (\bibinfo{date}{Aug.}
  \bibinfo{year}{2020}).
\newblock


\bibitem[\protect\citeauthoryear{Shneiderman and Muller}{Shneiderman and
  Muller}{2023}]%
        {ShneidermanM2023}
\bibfield{author}{\bibinfo{person}{Ben Shneiderman} {and}
  \bibinfo{person}{Michael Muller}.} \bibinfo{year}{2023}\natexlab{}.
\newblock \bibinfo{title}{On AI Anthropomorphism}.
\newblock
  \bibinfo{howpublished}{https://medium.com/human-centered-ai/on-ai-anthropomorphism-abff4cecc5ae}.
\newblock


\bibitem[\protect\citeauthoryear{Siddhartha}{Siddhartha}{2008}]%
        {siddhartha2008open}
\bibfield{author}{\bibinfo{person}{Siddhartha}.}
  \bibinfo{year}{2008}\natexlab{}.
\newblock \showarticletitle{Open-Source {H}induism}.
\newblock \bibinfo{journal}{\emph{Religion and the Arts}} \bibinfo{volume}{12},
  \bibinfo{number}{1-3} (\bibinfo{date}{Jan.} \bibinfo{year}{2008}),
  \bibinfo{pages}{34--41}.
\newblock


\bibitem[\protect\citeauthoryear{Sondarjee and Andrews}{Sondarjee and
  Andrews}{2022}]%
        {sondarjee2022decolonizing}
\bibfield{author}{\bibinfo{person}{Ma{\"\i}ka Sondarjee} {and}
  \bibinfo{person}{Nathan Andrews}.} \bibinfo{year}{2022}\natexlab{}.
\newblock \showarticletitle{Decolonizing International Relations and
  Development Studies: What’s in a Buzzword?}
\newblock \bibinfo{journal}{\emph{International Journal: Canada's Journal of
  Global Policy Analysis}} \bibinfo{volume}{77}, \bibinfo{number}{4}
  (\bibinfo{date}{Dec.} \bibinfo{year}{2022}), \bibinfo{pages}{551--571}.
\newblock


\bibitem[\protect\citeauthoryear{St.~Johns}{St.~Johns}{2023}]%
        {stjohns2023against}
\bibfield{author}{\bibinfo{person}{Mimi St.~Johns}.}
  \bibinfo{year}{2023}\natexlab{}.
\newblock \showarticletitle{Against Relationality: A Response to {A}beba
  {B}irhane}.
\newblock \bibinfo{journal}{\emph{GRACE: Global Review of AI Community Ethics}}
  \bibinfo{volume}{1}, \bibinfo{number}{1} (\bibinfo{date}{Feb.}
  \bibinfo{year}{2023}).
\newblock


\bibitem[\protect\citeauthoryear{Sun, Shen, Zhou, Zhang, Chen, Cox, Yang, and
  Gan}{Sun et~al\mbox{.}}{2023}]%
        {sun2023principle}
\bibfield{author}{\bibinfo{person}{Zhiqing Sun}, \bibinfo{person}{Yikang Shen},
  \bibinfo{person}{Qinhong Zhou}, \bibinfo{person}{Hongxin Zhang},
  \bibinfo{person}{Zhenfang Chen}, \bibinfo{person}{David Cox},
  \bibinfo{person}{Yiming Yang}, {and} \bibinfo{person}{Chuang Gan}.}
  \bibinfo{year}{2023}\natexlab{}.
\newblock \showarticletitle{Principle-Driven Self-Alignment of Language Models
  from Scratch with Minimal Human Supervision}.
\newblock In \bibinfo{booktitle}{\emph{Advances in Neural Information
  Processing Systems}}.
\newblock


\bibitem[\protect\citeauthoryear{Sundaram}{Sundaram}{2022}]%
        {sundaram2022hindutva}
\bibfield{author}{\bibinfo{person}{Dheepa Sundaram}.}
  \bibinfo{year}{2022}\natexlab{}.
\newblock \showarticletitle{{H}indutva 2.0: How a Conference on {H}indu
  Nationalism Launches a Change in Strategy for {N}orth {A}merican {H}indutva
  Organizations}.
\newblock \bibinfo{journal}{\emph{Journal of the American Academy of Religion}}
  \bibinfo{volume}{90}, \bibinfo{number}{4} (\bibinfo{year}{2022}),
  \bibinfo{pages}{809--814}.
\newblock


\bibitem[\protect\citeauthoryear{Swamy, Dann, Kidambi, Wu, and Agarwal}{Swamy
  et~al\mbox{.}}{2024}]%
        {swamy2024minimaximalist}
\bibfield{author}{\bibinfo{person}{Gokul Swamy}, \bibinfo{person}{Christoph
  Dann}, \bibinfo{person}{Rahul Kidambi}, \bibinfo{person}{Zhiwei~Steven Wu},
  {and} \bibinfo{person}{Alekh Agarwal}.} \bibinfo{year}{2024}\natexlab{}.
\newblock \bibinfo{title}{A Minimaximalist Approach to Reinforcement Learning
  from Human Feedback}.
\newblock \bibinfo{howpublished}{arXiv:2401.04056}.
\newblock


\bibitem[\protect\citeauthoryear{Tacheva and Ramasubramanian}{Tacheva and
  Ramasubramanian}{2023}]%
        {tacheva2023ai}
\bibfield{author}{\bibinfo{person}{Jasmina Tacheva} {and}
  \bibinfo{person}{Srividya Ramasubramanian}.} \bibinfo{year}{2023}\natexlab{}.
\newblock \showarticletitle{{AI} Empire: Unraveling the Interlocking Systems of
  Oppression in Generative {AI}'s Global Order}.
\newblock \bibinfo{journal}{\emph{Big Data \& Society}} \bibinfo{volume}{10},
  \bibinfo{number}{2} (\bibinfo{date}{Dec.} \bibinfo{year}{2023}),
  \bibinfo{pages}{20539517231219241}.
\newblock


\bibitem[\protect\citeauthoryear{Tharoor}{Tharoor}{2018}]%
        {tharoor2018hindu}
\bibfield{author}{\bibinfo{person}{Shashi Tharoor}.}
  \bibinfo{year}{2018}\natexlab{}.
\newblock \bibinfo{booktitle}{\emph{Why I Am a {H}indu}}.
\newblock \bibinfo{publisher}{Scribe}, \bibinfo{address}{Croydon, UK}.
\newblock


\bibitem[\protect\citeauthoryear{van Klinken}{van Klinken}{2020}]%
        {van2020studying}
\bibfield{author}{\bibinfo{person}{Adriaan van Klinken}.}
  \bibinfo{year}{2020}\natexlab{}.
\newblock \showarticletitle{Studying Religion in the Pluriversity: Decolonial
  Perspectives}.
\newblock \bibinfo{journal}{\emph{Religion}} \bibinfo{volume}{50},
  \bibinfo{number}{1} (\bibinfo{date}{Oct.} \bibinfo{year}{2020}),
  \bibinfo{pages}{148--155}.
\newblock


\bibitem[\protect\citeauthoryear{Varshney and Staggs}{Varshney and
  Staggs}{2024}]%
        {varshney2024hindu}
\bibfield{author}{\bibinfo{person}{Ashutosh Varshney} {and}
  \bibinfo{person}{Connor Staggs}.} \bibinfo{year}{2024}\natexlab{}.
\newblock \showarticletitle{{H}indu Nationalism and the New Jim Crow}.
\newblock \bibinfo{journal}{\emph{Journal of Democracy}} \bibinfo{volume}{35},
  \bibinfo{number}{1} (\bibinfo{date}{Jan.} \bibinfo{year}{2024}),
  \bibinfo{pages}{5--18}.
\newblock


\bibitem[\protect\citeauthoryear{Varshney}{Varshney}{2018}]%
        {Varshney2018}
\bibfield{author}{\bibinfo{person}{Kush~R. Varshney}.}
  \bibinfo{year}{2018}\natexlab{}.
\newblock \bibinfo{title}{{AI} and Safety}.
\newblock \bibinfo{howpublished}{Uehiro-Carnegie-Oxford Conference: Ethics and
  the Future of Artificial Intelligence}.
\newblock


\bibitem[\protect\citeauthoryear{Varshney, Keskar, and Socher}{Varshney
  et~al\mbox{.}}{2019}]%
        {varshney2019pretrained}
\bibfield{author}{\bibinfo{person}{Lav~R. Varshney},
  \bibinfo{person}{Nitish~Shirish Keskar}, {and} \bibinfo{person}{Richard
  Socher}.} \bibinfo{year}{2019}\natexlab{}.
\newblock \bibinfo{title}{Pretrained {AI} Models: Performativity, Mobility, and
  Change}.
\newblock \bibinfo{howpublished}{arXiv:1909.03290}.
\newblock


\bibitem[\protect\citeauthoryear{Viera~Magalh{\~a}es and
  Couldry}{Viera~Magalh{\~a}es and Couldry}{2021}]%
        {viera2021giving}
\bibfield{author}{\bibinfo{person}{Jo{\~a}o Viera~Magalh{\~a}es} {and}
  \bibinfo{person}{Nick Couldry}.} \bibinfo{year}{2021}\natexlab{}.
\newblock \showarticletitle{Giving by Taking Away: Big Tech, Data Colonialism
  and the Reconfiguration of Social Good}.
\newblock \bibinfo{journal}{\emph{International Journal of Communication}}
  \bibinfo{volume}{15} (\bibinfo{year}{2021}), \bibinfo{pages}{343--362}.
\newblock


\bibitem[\protect\citeauthoryear{Wang, Zhong, Li, Mi, Zeng, Huang, Shang,
  Jiang, and Liu}{Wang et~al\mbox{.}}{2023}]%
        {wang2023aligning}
\bibfield{author}{\bibinfo{person}{Yufei Wang}, \bibinfo{person}{Wanjun Zhong},
  \bibinfo{person}{Liangyou Li}, \bibinfo{person}{Fei Mi},
  \bibinfo{person}{Xingshan Zeng}, \bibinfo{person}{Wenyong Huang},
  \bibinfo{person}{Lifeng Shang}, \bibinfo{person}{Xin Jiang}, {and}
  \bibinfo{person}{Qun Liu}.} \bibinfo{year}{2023}\natexlab{}.
\newblock \bibinfo{title}{Aligning Large Language Models with Human: A Survey}.
\newblock \bibinfo{howpublished}{arXiv:2307.12966}.
\newblock


\bibitem[\protect\citeauthoryear{Weidinger, Uesato, Rauh, Griffin, Huang,
  Mellor, Glaese, Cheng, Balle, Kasirzadeh, Biles, Brown, Kenton, Hawkins,
  Stepleton, Birhane, Hendricks, Rimell, Isaac, Haas, Legassick, Irving, and
  Gabriel}{Weidinger et~al\mbox{.}}{2022}]%
        {weidinger2022taxonomy}
\bibfield{author}{\bibinfo{person}{Laura Weidinger}, \bibinfo{person}{Jonathan
  Uesato}, \bibinfo{person}{Maribeth Rauh}, \bibinfo{person}{Conor Griffin},
  \bibinfo{person}{Po-Sen Huang}, \bibinfo{person}{John Mellor},
  \bibinfo{person}{Amelia Glaese}, \bibinfo{person}{Myra Cheng},
  \bibinfo{person}{Borja Balle}, \bibinfo{person}{Atoosa Kasirzadeh},
  \bibinfo{person}{Courtney Biles}, \bibinfo{person}{Sasha Brown},
  \bibinfo{person}{Zac Kenton}, \bibinfo{person}{Will Hawkins},
  \bibinfo{person}{Tom Stepleton}, \bibinfo{person}{Abeba Birhane},
  \bibinfo{person}{Lisa~Anne Hendricks}, \bibinfo{person}{Laura Rimell},
  \bibinfo{person}{William Isaac}, \bibinfo{person}{Julia Haas},
  \bibinfo{person}{Sean Legassick}, \bibinfo{person}{Geoffrey Irving}, {and}
  \bibinfo{person}{Iason Gabriel}.} \bibinfo{year}{2022}\natexlab{}.
\newblock \showarticletitle{Taxonomy of Risks Posed by Language Models}. In
  \bibinfo{booktitle}{\emph{Proceedings of the ACM Conference on Fairness,
  Accountability, and Transparency}}. \bibinfo{pages}{214--229}.
\newblock


\bibitem[\protect\citeauthoryear{Widder, West, and Whittaker}{Widder
  et~al\mbox{.}}{2023}]%
        {widder2023open}
\bibfield{author}{\bibinfo{person}{David~Gray Widder}, \bibinfo{person}{Sarah
  West}, {and} \bibinfo{person}{Meredith Whittaker}.}
  \bibinfo{year}{2023}\natexlab{}.
\newblock \bibinfo{title}{Open (For Business): Big Tech, Concentrated Power,
  and the Political Economy of Open {AI}}.
\newblock \bibinfo{howpublished}{SSRN:4543807}.
\newblock


\bibitem[\protect\citeauthoryear{Wu, Hu, Shi, Dziri, Suhr, Ammanabrolu, Smith,
  Ostendorf, and Hajishirzi}{Wu et~al\mbox{.}}{2023}]%
        {Wu2023}
\bibfield{author}{\bibinfo{person}{Zeqiu Wu}, \bibinfo{person}{Yushi Hu},
  \bibinfo{person}{Weijia Shi}, \bibinfo{person}{Nouha Dziri},
  \bibinfo{person}{Alane Suhr}, \bibinfo{person}{Prithviraj Ammanabrolu},
  \bibinfo{person}{Noah~A. Smith}, \bibinfo{person}{Mari Ostendorf}, {and}
  \bibinfo{person}{Hannaneh Hajishirzi}.} \bibinfo{year}{2023}\natexlab{}.
\newblock \bibinfo{title}{Fine-Grained Human Feedback Gives Better Rewards for
  Language Model Training}.
\newblock \bibinfo{howpublished}{arXiv:2306.01693}.
\newblock


\bibitem[\protect\citeauthoryear{Zeng, Dai, Cheng, Hu, Chen, Du, and Xu}{Zeng
  et~al\mbox{.}}{2023}]%
        {zeng2023on}
\bibfield{author}{\bibinfo{person}{Dun Zeng}, \bibinfo{person}{Yong Dai},
  \bibinfo{person}{Pengyu Cheng}, \bibinfo{person}{Tianhao Hu},
  \bibinfo{person}{Wanshun Chen}, \bibinfo{person}{Nan Du}, {and}
  \bibinfo{person}{Zenglin Xu}.} \bibinfo{year}{2023}\natexlab{}.
\newblock \bibinfo{title}{On Diversified Preferences of Large Language Model
  Alignment}.
\newblock \bibinfo{howpublished}{arXiv:2312.07401}.
\newblock


\bibitem[\protect\citeauthoryear{Zheng, Chiang, Sheng, Zhuang, Wu, Zhuang, Lin,
  Li, Li, Xing, Zhang, Gonzalez, and Stoica}{Zheng et~al\mbox{.}}{2023}]%
        {zheng2023judging}
\bibfield{author}{\bibinfo{person}{Lianmin Zheng}, \bibinfo{person}{Wei-Lin
  Chiang}, \bibinfo{person}{Ying Sheng}, \bibinfo{person}{Siyuan Zhuang},
  \bibinfo{person}{Zhanghao Wu}, \bibinfo{person}{Yonghao Zhuang},
  \bibinfo{person}{Zi Lin}, \bibinfo{person}{Zhuohan Li},
  \bibinfo{person}{Dacheng Li}, \bibinfo{person}{Eric~P. Xing},
  \bibinfo{person}{Hao Zhang}, \bibinfo{person}{Joseph~E. Gonzalez}, {and}
  \bibinfo{person}{Ion Stoica}.} \bibinfo{year}{2023}\natexlab{}.
\newblock \bibinfo{title}{Judging {LLM}-as-a-Judge with {MT-B}ench and
  {C}hatbot {A}rena}.
\newblock \bibinfo{howpublished}{arXiv:2306.05685}.
\newblock


\bibitem[\protect\citeauthoryear{Zondi}{Zondi}{2020}]%
        {zondi2020decolonising}
\bibfield{author}{\bibinfo{person}{Siphamandla Zondi}.}
  \bibinfo{year}{2020}\natexlab{}.
\newblock \showarticletitle{Decolonising International Relations and Its
  Theory: A Critical Conceptual Meditation}.
\newblock In \bibinfo{booktitle}{\emph{Decolonisation after Democracy}},
  \bibfield{editor}{\bibinfo{person}{Laurence Piper}} (Ed.).
  \bibinfo{publisher}{Routledge}, \bibinfo{address}{London, UK},
  \bibinfo{pages}{16--31}.
\newblock


\end{thebibliography}

\end{document}